\documentclass[sigconf,screen,authorversion]{acmart}
\allowdisplaybreaks

\pdfoutput=1

\copyrightyear{2024} 
\acmYear{2024} 
\setcopyright{acmlicensed}\acmConference[LICS '24]{39th Annual ACM/IEEE Symposium on Logic in Computer Science}{July 8--11, 2024}{Tallinn, Estonia}
\acmBooktitle{39th Annual ACM/IEEE Symposium on Logic in Computer Science (LICS '24), July 8--11, 2024, Tallinn, Estonia}
\acmDOI{10.1145A/3661814.3662131}
\acmISBN{979-8-4007-0660-8/24/07}

\usepackage{makecell}
\usepackage[T1]{fontenc}
\usepackage{graphicx}
\usepackage[all]{xy}
\usepackage{amsmath, amsfonts}
\usepackage{tikz-cd}
\usepackage{todonotes}
\usepackage{mathtools}
\usepackage{mathrsfs}  
\usepackage{float}
\usepackage{pgfplots}
\usetikzlibrary{backgrounds}
\tikzstyle{background rectangle}=  [fill=gray!10]
\tikzstyle{background rectangle 2}=  [fill=olive!10]
\tikzstyle{background rectangle 3}=  [fill=teal!8]

\usepackage{hyperref}
\pgfplotsset{compat=1.17}
\usepgfplotslibrary{colormaps}
\usepgfplotslibrary{fillbetween}
\usepackage{xcolor}
\definecolor{myred}{RGB}{255, 82, 82}
\definecolor{mylightred}{RGB}{255, 126, 126}
\definecolor{myblue}{RGB}{52, 152, 219}
\definecolor{mylightblue}{RGB}{93, 173, 226}
\definecolor{mygreen}{RGB}{46, 204, 113}
\definecolor{mylightgreen}{RGB}{88, 214, 141}
\usepackage{subcaption}
\usepackage{thm-restate}

\theoremstyle{definition}
\newtheorem{notation}{Notation}

\usetikzlibrary{calc}
\usetikzlibrary{decorations.pathmorphing}

\tikzset{curve/.style={settings={#1},to path={(\tikztostart)
    .. controls ($(\tikztostart)!\pv{pos}!(\tikztotarget)!\pv{height}!270:(\tikztotarget)$)
    and ($(\tikztostart)!1-\pv{pos}!(\tikztotarget)!\pv{height}!270:(\tikztotarget)$)
    .. (\tikztotarget)\tikztonodes}},
    settings/.code={\tikzset{quiver/.cd,#1}
        \def\pv##1{\pgfkeysvalueof{/tikz/quiver/##1}}},
      quiver/.cd,pos/.initial=0.35,height/.initial=0}
    
\newcommand{\N}{{\mathbb N}}
\newcommand{\Ngz}{\N_{\scalebox{0.7}{$>\!\!0$}}}
\newcommand{\R}{{\mathbb R}}

\newcommand{\G}{{G}} %
\newcommand{\Gsub}{\G_{\scalebox{0.6}{$\!\leq$}}}
\newcommand{\Meas}{\mathbf{Meas}}

\newcommand{\Sbs}{\mathbf{Sbs}}
\newcommand{\supp}{\mathrm{supp}}
\newcommand{\pairing}[2]{\left\langle #1, #2 \right\rangle}
\newcommand{\ev}{\mathrm{ev}}
\newcommand{\mc}{\textsl{mc}}
\newcommand{\ord}{\textsl{ord}}
\newcommand{\id}{\mathrm{id}}
\newcommand{\Aut}{\mathrm{Aut}}
\newcommand{\mn}{\textsl{mn}}
\newcommand{\pmn}{\textsl{pmn}}

\newcommand{\elemfreedisc}{{\elemfree_{\!\textsl{d}}}}
\newcommand{\elemfree}{\nabla}
\newcommand{\diff}{\mathrm{d}}

\newcommand{\Part}{\mathcal{P}}

\newcommand{\total}{\mathrm{tt}}
\newcommand{\atoms}{\mathcal{A}}
\newcommand{\iid}{\textsl{iid}}

\newcommand{\eg}{\emph{e.g.}}
\newcommand{\ie}{\emph{i.e.}}

\newcommand{\One}{\mathbf{1}}

\newcommand{\base}{\textsl{base}}
\newcommand{\uniform}{\mathscr{U}_{[0, 1]}}
\newcommand{\Kl}{\mathcal{K}\ell}
\newcommand{\dd}{\textsl{dd}}
\newcommand{\pdd}{\textsl{pdd}}

\DeclarePairedDelimiter{\size}{\parallel}{\parallel}

\usepackage{bm}

\newsavebox\sbpto
\savebox\sbpto{\begin{tikzpicture}[baseline=-2.5pt]
            \node [draw, shape = circle, fill = white, inner sep=0.8pt] (0,0){};
                \end{tikzpicture}}
\newcommand{\kerto}{\mathrel{\ooalign{$\rightarrow$\cr\hfil\!$\usebox\sbpto$\hfil\cr}}}

\newcommand{\leftmulti}{\bm{(}\hspace{-4.2pt}\bm{(}}
\newcommand{\rightmulti}{\bm{)}\hspace{-4.2pt}\bm{)}}
\newcommand{\multicoeff}[1]{\leftmulti #1 \rightmulti}
\newcommand{\multicoeffpart}[1]{\leftmulti #1 \rightmulti_\mathrm{p}}
\newcommand{\M}{\mathcal{M}}

\usetikzlibrary{decorations.markings}
\tikzset{ker/.style={/utils/exec=\tikzset{every node/.append style={outer sep=0ex}},
postaction=decorate,decoration={markings, mark=at position 0.5 with {\node [draw, shape = circle, fill = white, inner sep=0.8pt] (0,#1){};}}},
ker/.default=0.75ex}

\begin{document}

\title{Element-free probability distributions and random partitions}

\author{Victor Blanchi}
\email{victor.blanchi@ens.psl.eu}
\affiliation{
  \institution{\'Ecole Normale Sup\'erieure, Universit\'e PSL}
  \institution{and IRIF, Universit\'e Paris Cit\'e}
  \city{Paris}
  \country{France}
}
\author{Hugo Paquet}
\email{paquet@lipn.fr}
\affiliation{%
  \institution{Laboratoire d'Informatique de Paris Nord}
  \city{Villetaneuse}
  \country{France}
}




\begin{abstract}
  An ``element-free'' probability distribution is what remains of a
  probability distribution after we forget the elements to which the
  probabilities were assigned. These objects naturally arise in
  Bayesian statistics, in situations where elements are used as
  labels and their specific identity is not important.
  
  This paper develops the structural theory of element-free
  distributions, using multisets and category theory. We give 
  operations for moving between element-free and ordinary distributions, and
  we show that these operations commute with multinomial sampling.
  
  We then exploit this theory to prove two representation theorems.
  These theorems show that element-free distributions provide a natural
  representation for key random structures in Bayesian nonparametric
  clustering: exchangeable random partitions, and random distributions
  parametrized by a base measure.
  
\end{abstract}

\maketitle

\section{Introduction}

This paper is about the idea of ``element-free'' probability
theory. We can take any probability distribution and forget the
elements to which probabilities have been assigned, keeping only the
probability coefficients. For instance, given the distribution on the
set $\{ a, b, c \}$ assigning $\frac{1}{5}$ to $a$ and $b$, and $\frac{3}{5}$ to
$c$, if we forget the elements $a$, $b$, and $c$, we are left with coefficients $\frac{1}{5}, \frac{1}{5},
\frac{3}{5}$. These coefficients are unordered and form a multiset (or
bag) written $[\frac{1}{5}, \frac{1}{5}, \frac{3}{5}]$. We call this
kind of multiset an element-free distribution (\S\ref{sec:discr-elem-free}).

Clearly, the element-free approach involves a significant loss of
information. But this is justified in situations where the precise
identity of elements is not important. Our main motivation comes from
nonparametric Bayesian statistics: element-free distributions
are deeply connected to the theory of random exchangeable partitions
(\eg~the Chinese Restaurant Process \cite[\emph{p.} 92]{10.1007/BFb0099421})
and random probability distributions parametrized by a base measure
(\eg~the Dirichlet process \cite{ferguson1973bayesian}). 

In this paper we develop the theory of element-free
distributions. We also explain their canonical status and clarify the connection to
partitions and random distributions. We sketch the main ideas in this
introduction.

\subsection{Element-free sampling and random partitions}
\label{sec:rand-part-bayes}

We first explain the connection between element-free probability and
partitions. The point is that partitions are the element-free
counterpart of lists of elements.

Given a list $(x_1, \dots, x_K) \in X^K$, where $X$ is any set and $K$
a positive integer, we can construct a partition of the set
$\{ 1, \dots, K\}$ in which $i$ and $j$ are in the same block whenever
$x_i = x_j$. For instance, taking $X = \{ a, b, c\}$, the list $(a, a, b, a, c)$
induces the partition $\{ \{ 1, 2, 4 \} , \{3 \}, \{ 5\}\}$. This partition is an element-free abstraction of the list:
the elements themselves have been forgotten but we record the indices
at which equal elements appeared.

A key operation in probability and statistics is that of drawing
lists of independent and identically distributed (\emph{iid}) samples according
to a probability distribution. There is also an element-free version of
iid sampling, for generating a partition directly from an
element-free distribution \cite{lics-jacobs}. The two operations of forgetting elements
and iid sampling commute:
  
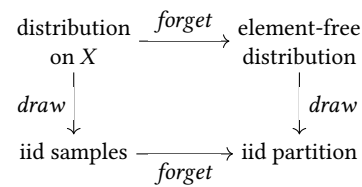
\[
\begin{tikzpicture}
\node[align=center] (dist) at (0, 1.5) {distribution\\ on $X$};
\node[align=center] (elfree) at (3, 1.5) {element-free\\ distribution};
\node (samples) at (0, 0) {iid samples};
\node (part) at (3, 0) {iid partition};

\draw[->] (samples) -- (part) node [midway, below, fill=white] {\textit{forget}};
\draw[->] (dist) -- (samples) node [midway, left, fill=white] {\textit{draw}};
\draw[->] (elfree) -- (part) node [midway, right, fill=white] {\textit{draw}};
\draw[->] (dist) -- (elfree) node [midway, above, fill=white] {\textit{forget}};

   \end{tikzpicture}
    \]
    A key contribution of this paper is to take this idea further, to
    incorporate continuous distributions on measurable spaces. 
    In the continuous setting we can prove a new representation theorem
    for random partitions (\S\ref{sec:exch-mult}, \S\ref{sec:random-element-free}).

\subsection{Motivation from Bayesian clustering}
To motivate this work we explain the connection to statistical practice. Our purpose is to clarify the foundations, and we
do not propose any new methods. (We briefly discuss
connections to probabilistic programming
in \S\ref{sec:conclusion}.)

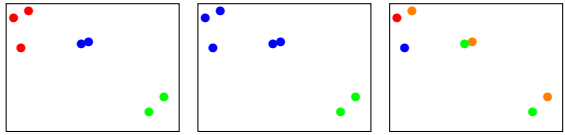
\begin{figure}[t]
\begin{tikzpicture}
\node (a) at (0, 2) {\color{red} $\bullet$};
\node (a) at (0.1, 1.6) {\color{red} $\bullet$};
\node (a) at (0.2, 2.1) {\color{red} $\bullet$};
\node (a) at (1, 1.68) {\color{blue} $\bullet$};
\node (a) at (0.9, 1.65) {\color{blue} $\bullet$};
\node (a) at (2, 0.95) {\color{green} $\bullet$};
\node (a) at (1.8, 0.75) {\color{green} $\bullet$};
\draw (-0.1, 0.5) -- (-0.1, 2.2) -- (2.2, 2.2) -- (2.2, 0.5) -- (-0.1, 0.5);
\end{tikzpicture}
\   
\begin{tikzpicture}
\node (a) at (0, 2) {\color{blue} $\bullet$};
\node (a) at (0.1, 1.6) {\color{blue} $\bullet$};
\node (a) at (0.2, 2.1) {\color{blue} $\bullet$};
\node (a) at (1, 1.68) {\color{blue} $\bullet$};
\node (a) at (0.9, 1.65) {\color{blue} $\bullet$};
\node (a) at (2, 0.95) {\color{green} $\bullet$};
\node (a) at (1.8, 0.75) {\color{green} $\bullet$};
\draw (-0.1, 0.5) -- (-0.1, 2.2) -- (2.2, 2.2) -- (2.2, 0.5) -- (-0.1, 0.5);
\end{tikzpicture}
\ 
\begin{tikzpicture}
\node (a) at (0, 2) {\color{red} $\bullet$};
\node (a) at (0.1, 1.6) {\color{blue} $\bullet$};
\node (a) at (0.2, 2.1) {\color{orange} $\bullet$};
\node (a) at (1, 1.68) {\color{orange} $\bullet$};
\node (a) at (0.9, 1.65) {\color{green} $\bullet$};
\node (a) at (2, 0.95) {\color{orange} $\bullet$};
\node (a) at (1.8, 0.75) {\color{green} $\bullet$};
\draw (-0.1, 0.5) -- (-0.1, 2.2) -- (2.2, 2.2) -- (2.2, 0.5) -- (-0.1, 0.5);
\end{tikzpicture}
\caption{Three different clusterings for a set of points, with
  clusters represented as colours.}
  \label{fig:clustering-example}
\end{figure}

In data analysis, clustering is the problem of appropriately
partitioning a dataset $y_1, \dots, y_K$ into clusters (Figure~\ref{fig:clustering-example}). In the Bayesian approach to
clustering, the goal is not to find the ``best'' partition, but to
find a suitable probability distribution over
partitions. In \emph{nonparametric} clustering the total number of
clusters is not fixed, and grows as more data is included. A Bayesian
model should first be specified independently of the data, and so
should only be concerned with partitions of the indexing set $\{1,
\dots, K\}$.

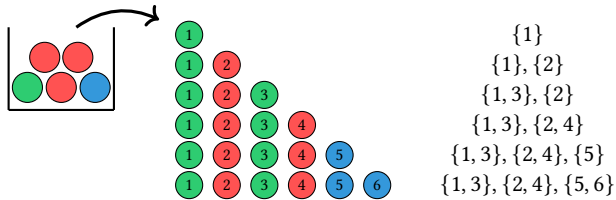
\begin{figure}[t]
\begin{tikzpicture}
    \draw[thick] (-0.7,1.1) -- (-0.7,0); 
    \draw[thick] (-0.7,0) -- (0.7,0); 
    \draw[thick] (0.7,0) -- (0.7,1.1);  
    
    \draw [fill = myred] (0, 0.3) circle (0.2); 
    \draw [fill = mygreen] (-0.45,0.3) circle (0.2); 
    \draw [fill = myblue] (0.45,0.3) circle (0.2); 
    \draw [fill = myred] (0.22, 0.7) circle (0.2); 
    \draw [fill = myred] (-0.22, 0.7) circle (0.2);

    \draw[->, very thick] (0.2, 1.1) to[bend left=30] (1.3,
    1.2);

    \draw [fill=mygreen] (1.7, 1) circle (0.18);
    \node at (1.7, 1) {\scriptsize $1$}; 

    \draw [fill=mygreen] (1.7, 0.6) circle (0.18);
    \node at (1.7, 0.6) {\scriptsize $1$}; 

    \draw [fill=myred] (2.2, 0.6) circle (0.18);
    \node at (2.2, 0.6) {\scriptsize $2$}; 

    \draw [fill=mygreen] (1.7, 0.2) circle (0.18);
    \node at (1.7, 0.2) {\scriptsize $1$}; 

    \draw [fill=myred] (2.2, 0.2) circle (0.18);
    \node at (2.2, 0.2) {\scriptsize $2$}; 
    
    \draw [fill=mygreen] (2.7, 0.2) circle (0.18);
    \node at (2.7, 0.2) {\scriptsize $3$}; 

    \draw [fill=mygreen] (1.7, -0.2) circle (0.18);
    \node at (1.7,- 0.2) {\scriptsize $1$}; 

    \draw [fill=myred] (2.2,- 0.2) circle (0.18);
    \node at (2.2,- 0.2) {\scriptsize $2$}; 
    
    \draw [fill=mygreen] (2.7, -0.2) circle (0.18);
    \node at (2.7, -0.2) {\scriptsize $3$}; 

    \draw [fill=myred](3.2, -0.2) circle (0.18);
    \node at (3.2, -0.2) {\scriptsize $4$}; 

    \draw [fill=mygreen] (1.7, -0.6) circle (0.18);
    \node at (1.7,- 0.6) {\scriptsize $1$}; 

    \draw [fill=myred] (2.2,- 0.6) circle (0.18);
    \node at (2.2,- 0.6) {\scriptsize $2$}; 
    
    \draw [fill=mygreen] (2.7, -0.6) circle (0.18);
    \node at (2.7, -0.6) {\scriptsize $3$}; 

    \draw [fill=myred](3.2, -0.6) circle (0.18);
    \node at (3.2, -0.6) {\scriptsize $4$};

    \draw [fill=myblue](3.7, -0.6) circle (0.18);
    \node at (3.7,  -0.6) {\scriptsize $5$}; 

        \draw [fill=mygreen] (1.7, -1) circle (0.18);
    \node at (1.7,- 1) {\scriptsize $1$}; 

    \draw [fill=myred] (2.2, -1) circle (0.18);
    \node at (2.2,- 1) {\scriptsize $2$}; 
    
    \draw [fill=mygreen] (2.7, -1) circle (0.18);
    \node at (2.7, -1) {\scriptsize $3$}; 

    \draw [fill=myred](3.2, -1) circle (0.18);
    \node at (3.2, -1) {\scriptsize $4$};

    \draw [fill=myblue](3.7, -1) circle (0.18);
    \node at (3.7, -1) {\scriptsize $5$}; 

    \draw [fill=myblue](4.2, -1) circle (0.18);
    \node at (4.2, -1) {\scriptsize $6$};

    \node at (6.2, 1) {$ \{ 1  \}$};
    \node at (6.2, 0.6) {$ \{ 1\}, \{ 2\} $};
    \node at (6.2, 0.2) {$ \{ 1, 3\}, \{ 2\} $};
    \node at (6.2, -0.2) {$ \{ 1, 3\}, \{ 2, 4\} $};
    \node at (6.2, -0.6) {$ \{ 1, 3\}, \{ 2, 4\}, \{ 5\} $};
    \node at (6.2, -1) {$ \{ 1, 3\}, \{ 2, 4\}, \{ 5, 6\} $};
      \end{tikzpicture}
\caption{Iteratively sampling a partition from a probability distribution
  represented as an urn. (Here \emph{forgetting elements} corresponds
  to  forgetting the colours.)}
  \label{fig:urn}
\end{figure}

To generate a random partition, a popular method is to fix a probability distribution
over a space whose elements will be used as cluster names, and then
sample a name for each index $k\leq K$.
This method is illustrated in Figure~\ref{fig:urn} using an urn of coloured
balls. Observe how new clusters are created as $k$ grows.

The specific names (or colours) used for clusters are not important
for the partition model: substituting red balls for balls of a new colour would
not affect the resulting distribution on partitions. In other words,
it is only the element-free distribution that matters. 

It makes sense to study this particular clustering method in depth,
because it is universal: all well-behaved distributions on partitions
can be generated in this way from a \emph{random} element-free
distribution. We discuss this further in \S\ref{sec:exch-mult}.

\subsection{Base measures and random distributions}
\label{sec:base-measures-random}

Another key contribution of this paper is to show that, as well as
forgetting elements, we can reconstruct them in a principled way.
The idea is to sample them from a fixed distribution called a base
measure. We show that, if $\mu$ is a distribution on a space $X$, we have a
reverse situation:
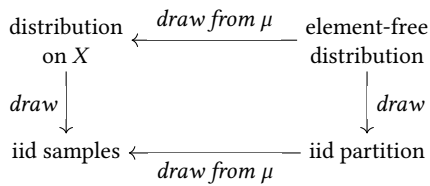
\[
  \begin{tikzpicture}
\node[align=center] (dist) at (0, 1.5) {distribution\\ on $X$};
\node[align=center] (elfree) at (4, 1.5) {element-free\\ distribution};
\node (samples) at (0, 0) {iid samples};
\node (part) at (4, 0) {iid partition};

\draw[->]  (part) -- (samples)  node [midway, below, fill=white]
{\textit{draw from $\mu$}};
\draw[->] (dist) -- (samples) node [midway, left, fill=white] {\textit{draw}};
\draw[->] (elfree) -- (part) node [midway, right, fill=white] {\textit{draw}};
\draw[->]  (elfree) -- (dist) node [midway, above, fill=white] {\textit{draw from $\mu$}};

   \end{tikzpicture}
\]
These operations are all probabilistic. Given an element-free distribution, our construction does not give a
fixed distribution on $X$, but a random distribution: a distribution
on distributions.

The use of a base measure is common in nonparametric
statistics. Indeed a number of key models for random distributions
(\eg~the Dirichlet or Pitman-Yor processes \cite{ferguson1973bayesian,
  pitman1997two}) admit a base measure as a
parameter, precisely because these models are only
concerned with the element-free part of the distributions they generate. (The popular \emph{stick-breaking}
methods \cite{sethuraman1994constructive} can be regarded as generating element-free distributions.)

In \S\ref{sec:random-element-free} we study random
distributions parametrized by a base measure, and give a new
representation theorem (Theorem~\ref{thm:correspondence}): the parametrized random distributions
corresponding to random element-free distributions are in bijection
with certain natural transformations between functors of
distributions. This appears to be a new perspective on random
distributions. 

\subsection{Multisets and exchangeability}
\label{sec:exch-mult}

This paper is heavily based on
multisets. The first reason is that the coefficients in an
element-free distribution are unordered, just like the elements to which they
were originally assigned.

The second reason is that
distributions on multisets correspond to distributions on lists that
are invariant under list permutation. This kind of permutation-invariance, known as exchangeability, is
a fundamental notion in Bayesian modelling, because the
way we have indexed the data points $y_1, \dots, y_K$ is
usually arbitrary (\cite[\S1.2]{gelman1995bayesian}). By working with
multisets, we implicitly enforce exchangeability of all distributions. 

For partitions, exchangeability means that, for instance,
partitions $\{\{1, 2\}, \{3\}\}$ and $\{\{1, 3\},\{2\}\}$ have the
same probability of occurring. The model of Figure~\ref{fig:urn} is exchangeable
because the colours are sampled independently and from the same urn.

\paragraph{Multinomial distributions.} The multinomial construction
(\S\ref{sec:sampl-with-elem}) is
a function from distributions on $X$ to distributions on multisets
over $X$. This is a model for the process of drawing a list of iid
samples and then forgetting the list order. 

To enforce exchangeability in the element-free setting, we
use \emph{integer partitions} rather than the usual set partitions. An
integer partition of $K$ (see also \S\ref{sec:multisets-partitions}) is a multiset of positive integers which sum
to $K$, representing the ``block sizes'' in a partition of the set
$\{ 1, \dots, K\}$.  The 
element-free version of multinomial sampling is a function from element-free
distributions to distributions over integer partitions (\S\ref{sec:sampl-without-elem}).

We emphasize that integer partitions are the element-free counterpart
of multisets of elements: for example the multiset $[a, a, a, b, c]$
gives rise to the partition $[3, 1, 1]$ of the integer $5$. (There is
a beautiful mathematical theory relating lists, multisets, set
partitions and integer partitions \cite{DBLP:conf/csl/0001S23,jacobsone}.)  In
the rest of this paper, by \emph{partition} we always mean integer
partition.

\paragraph{Kingman's theorem.}
A central contribution of this paper is a new categorical proof of
Kingman's representation theorem for random partitions. This theorem
states that, in the limit $K = \infty$, every exchangeable random partition
is induced by a unique random element-free distribution via multinomial
sampling (as in Figure~\ref{fig:urn}).

Kingman's theorem is the element-free counterpart of de Finetti's
theorem \cite{de1929function}, a foundational result in probability theory,
that has recently been given various categorical presentations \cite{fritz2021finetti,samnote,sam-bart}.  This
paper makes clear the connection between the two theorems, and we can
derive Kingman's theorem using a purely categorical argument
(Theorem~\ref{thm:kingman}), exploiting
the results of \S\ref{sec:general-element-free} and \S\ref{sec:sampl-mult-part}
connecting element-free and ordinary probability theory. We emphasize that continuous
distributions, and thus measure theory, are essential for this theorem
to hold: discrete element-free distributions \emph{do not} represent
all exchangeable random partitions \cite{kingman-representation}.

\subsection{Summary of contributions and outline}

We summarize the key contributions of this paper.

In \S\ref{sec:general-element-free}, we construct a measurable space of element-free
  distributions defined as infinite multisets. We define two fundamental
  operations: \emph{multiplicity count} for forgetting elements (following \cite{lics-jacobs}), and drawing elements from a base
  measure.

  In \S\ref{sec:multisets-partitions} we consider spaces of
    finite multisets and partitions, and also define multiplicity
    count \cite{lics-jacobs} and base measures. We then make formal,
    in \S\ref{sec:sampl-mult-part}, the informal
    diagrams of \S\ref{sec:rand-part-bayes} and
    \S\ref{sec:base-measures-random}: firstly by defining an
    element-free version of multinomial sampling for partitions,
    and secondly by showing that multinomial sampling commutes with the two
    fundamental operations (multiplicity count, and drawing from a base measure). 

  We can then prove our two main theorems. In \S\ref{sec:kingman} we give a new statement and proof of
    Kingman's theorem in terms of a categorical limit, using the
    categorical version of de Finetti's theorem for standard Borel
    spaces \cite{samnote}. In \S\ref{sec:random-element-free} we prove a new
    representation theorem for random distributions parametrized by a
    base measure, in terms of a correspondence between natural
    transformations and random element-free distributions.

At a high level our motivation is to establish the following
dictiorary as a tool for reasoning about models for clustering.
\[
  \begin{tabular}{|c|c|}
  \hline
  \textsc{Elements in space $X$} &   \textsc{Element-free} \\
  \hline
  \textcolor{gray} {Lists $\langle x_1, \dots, x_K\rangle$ }  & \textcolor{gray}  {Partitions of the set $\{ 1, \dots, K\}$ } \\
  Multisets $[ x_1, \dots, x_K]$ &   Partitions of the number $K$ \\
  Distributions on $X$ &   Element-free distributions \\
  de Finetti's theorem &   Kingman's theorem\\
  \hline
\end{tabular}
\]

We begin in \S\ref{sec:prob-theory-atoms} with a recap section on
measure-theoretic probability theory. Then in
\S\ref{sec:discr-elem-free} we introduce element-free distributions
via the simple special case of distributions on $\N$.

\section{Probability theory and atoms}
\label{sec:prob-theory-atoms}
This section contains preliminary background on measure theory, the
notion of probability kernel, and atoms.

A probability distribution on a set $X$ is a function $p : X \to [0,
1]$ such that $\sum_{x \in X} p(x) = 1$. This definition works well
for discrete probability, but for continuous probability we need
measure theory. 

\begin{definition}
A measurable space is a set $X$ equipped with a $\sigma$-algebra
$\Sigma_X$: this is a set of subsets of $X$, containing $\emptyset$
and closed under complements, and countable unions and intersections.
\end{definition}
Elements of $\Sigma_X$ are called measurable subsets of $X$.  We often refer to the space as $X$, omitting the $\Sigma_X$, when there is no ambiguity. Key examples of measurable spaces include: 
  \begin{itemize}
  \item The set $\R$ of real numbers equipped with the Borel sets
    $\Sigma_\R$. This is defined as the smallest  $\sigma$-algebra containing all intervals $(a, b) \subseteq \R$.
  \item Any subset of $S \subseteq \R$, with the $\sigma$-algebra $\{
    S \cap U \mid U \in \Sigma_\R \}$.
   \item Any set $X$ with the discrete $\sigma$-algebra containing all
     subsets of $X$. With discrete spaces we can view 
     distributions on sets as instances of a general notion of
     probability measure. 
  \end{itemize}

  \begin{definition}
A probability measure on a measurable space $(X, \Sigma_X)$ is a
function $\mu : \Sigma_X \to [0, 1]$ such that $\mu(X) = 1$, and for
any countable family of disjoint subsets $U_i \in \Sigma_X$, we have
$\mu( \bigcup_i U_i) = \sum_{i} \mu(U_i).$ 
\end{definition}

There is a category $\Meas$ of measurable spaces, whose morphisms $(X,
\Sigma_X) \to (Y, \Sigma_Y)$ are measurable functions, i.e. maps $f :
X \to Y$ such that for every $U \in \Sigma_Y$, $f^{-1}U \in
\Sigma_X$.

We can pushforward a probability measure $\mu$
on $X$ along a measurable function, and get a probability measure $f_*\mu$ on
$Y$ given by $(f_*\mu)(U) = \mu(f^{-1}U)$.

For any measurable function $f : X \to \R_+$ and probability measure
$\mu$ on $X$, we can define its Lebesgue integral $\int_{x \in X} f(x)
\mu(\diff x)
\in \R_+ \cup \{ \infty \}$.

\subsection{Kernels and the Giry monad}
There is a monad of probability distributions on $\Meas$,
traditionally called the Giry monad \cite{giry, lawvere1963functorial}. For a measurable space $X$, we write $\G X$ for the set of probability
  measures on $X$, equipped with the smallest $\sigma$-algebra which
  makes the  maps $\ev_U : \G X \to [0, 1] : \mu \mapsto \mu(U)$
  measurable for every $U \in \Sigma_X$. 
  For a measurable function $f : X \to Y$, there is a measurable function $\G f
  : \G X \to \G Y$ given by pushforward along $f$. This
  defines a functor $\G :  \Meas \to \Meas$.

  \begin{definition}
  A \emph{(probability) kernel} from $X$ to $Y$ is a measurable function
  $X \to \G Y$. We use the notation $X \kerto Y$.
\end{definition}
  
  A kernel $k : X \kerto Y$ lifts to a function $k^\dagger :
  \G X \to \G Y$ given by $\mu \mapsto \int_{x \in X} k(x, -)
  \mu(\diff x)$, and we can use this to compose kernels: if $h : Y \kerto
  Z$, then $h^\dagger \circ k : X \to \G Z$ is a kernel $X \kerto 
  Z$. 
  
  Finally, for every space $X$ there is a map $\eta_X : X \to \G X$ which maps
  $x \in X$ to the Dirac distribution $\delta_x : U \mapsto [x \in U]$,
  which returns $x$ with probability $1$. 

  \begin{lemma}
 The triple $(\G, (-)^\dagger, \eta)$ determines a monad on $\Meas$.
  \end{lemma}

  \paragraph{Subprobability distributions.} A subprobability measure
  on $(X, \Sigma_X)$ is defined in the same way as a probability measure, only with
  the weaker requirement that $\mu(X) \leq 1$. There is 
  a space $\Gsub X$ of subprobability measures, that contains $\G X$ as a
  subspace; the definitions are very similar and we omit the details. ($\Gsub$ is also a monad, but we do not use this.)
  
  \subsection{Atoms and standard Borel spaces}

For this paper we restrict ourselves to a class of well-behaved
spaces (e.g.~\cite[\emph{p.~}11]{cinlar2011probability}) in which it is safe to reason about elements and atoms. 
  
\begin{definition}
A measurable space is a \emph{standard Borel space} if it is either
discrete and countable, or measurably isomorphic to $(\R, \Sigma_\R)$. 
\end{definition}

Standard Borel spaces include any measurable subset of $(\R, \Sigma_\R)$,
and they are closed under countable products. The Giry monad restricts to
the subcategory $\Sbs$ of standard Borel spaces, i.e.~if $X$ is standard
Borel then so is $\G X$. In this paper when we write $\Kl(\G)$ we mean
the Kleisli category for $\G$ over $\Sbs$. This is the category whose
objects are standard Borel spaces and whose morphisms are kernels.

In a standard Borel space, all singletons are measurable. If $X \in
\Sbs$, and $\mu \in \G X$ then we say $\mu$ has an \emph{atom} at $x
\in X$ if $\mu\{ x\} > 0$. The set of atoms of $\mu$, written $\atoms_\mu$, must be
countable and thus a measurable subset of $X$. Say $\mu$ is
discrete or \emph{purely atomic} if $\mu(X) = \mu(\atoms_\mu)$, and
\emph{non-atomic} if $\atoms_\mu = \emptyset$.  

\paragraph{Product spaces and $\iid$ measures.} The product of measurable
spaces $(X, \Sigma_X)$ and $(Y, \Sigma_Y)$ is the set $X \times Y$
equipped with the $\sigma$-algebra generated by the rectangles $U
\times V$ ($U \in \Sigma_X$, $V \in \Sigma_Y$). 

Probability measures $\mu \in \G X$ and $\nu \in \G Y$ determine a
unique product measure $\mu \times \nu$ which satisfies $(\mu \times
\nu)(U \times V) = \mu(U) \times \nu(V)$.
In particular, for any $\kappa \in \N \cup \{ \infty\}$, there is a
measurable function $\iid_{\kappa} : GX \to G(X^\kappa)$ which sends $\mu$
to the $\kappa$-fold product measure $\mu^\kappa$. 

\section{Discrete element-free distributions}
\label{sec:discr-elem-free}

This section is about a simple special case:  probability
distributions over the natural numbers $\N$, which are always
discrete. For every distribution on $\N$, there is an induced element-free
distribution: this is the family of probability
coefficients, without the elements to which they were assigned.

We define discrete element-free distributions as multisets of values
in $(0, 1]$ which sum to $1$. We formalize multisets as functions
$(0, 1]\to \N$, indicating the multiplicity of each value. 
These multisets may be (countably) infinite, because distributions
on $\N$ can have infinite support.
\begin{definition}
  A \emph{discrete element-free distribution} is a function $\varphi :
  (0, 1] \to \N$ whose support $\supp(\varphi) = \{r \in (0, 1] \mid \varphi(r) > 0
  \}$ is (finite or) countable, and such that $\sum_{r \in \supp(\varphi)} \varphi(r)
  \cdot r = 1$. 
\end{definition}
(Note that, although the multisets may be infinite, each $r \in (0, 1]$
must have a finite multiplicity.)

\paragraph{A space of element-free distributions.}

The set of all discrete element-free
distributions is denoted $\elemfreedisc$. It is equipped with the $\sigma$-algebra generated by the sets
$E_{k}^U = \{ \varphi \in \elemfreedisc \mid \textstyle \sum_{r \in U \cap \supp(\varphi)} \varphi(r)= k  \} 
$
for $k \in \N$ and $U \in \Sigma_{(0, 1]}$. Informally, $E^U_k$ is the set
of multisets having exactly $k$ elements in $U$. (There is also a measurable set $E^U_\infty := \elemfreedisc \setminus \bigcup_{k \in \N} E^U_K$.)

We note that this is the smallest 
$\sigma$-algebra with respect to which the maps
$
\ev_U : \elemfreedisc \to \N \cup \{ \infty \}$, 
where $\ev_U(\varphi) = \sum_{r \in U \cap \supp(\varphi)} \varphi(r)
$, are measurable. (The similarity with the $\sigma$-algebra on
$\G(0, 1]$ is not an accident: multisets are often formalized as
integer-valued measures in the probability literature (\eg~\cite{last-penrose}).)

\paragraph{Multiplicity count.}
Every distribution on $\N$ has an underlying discrete element-free
distribution. This is computed by an operation called \emph{multiplicity
  count} (because it counts the multiplicities of coefficients
\eg~\cite{lics-jacobs}). We show that this is a measurable function.

\begin{lemma}
  \label{lem:mc-disc-def}
  Every distribution $\mu \in \G\N$ induces a \emph{multiplicity count} $\mc(\mu) \in \elemfreedisc$ given by
  \[ \mc(\mu)(r) = \#\{ n \in \N \mid \mu\{n\}= r \}\]
  for $r \in (0, 1]$. The function $\mc : \G\N \to \elemfreedisc$ is
  measurable. 
\end{lemma}
\begin{proof}
The preimage of a generating set $E_k^U \in \Sigma_\elemfreedisc$, where $k
\in \N$ and $U \in \Sigma_{(0,1]}$, under the function $\mc$, is the
set of distributions $\mu$ such that $\mu\{n_i\} \in U$ for precisely
$k$ values $n_1, \dots, n_k \in \N$. Thus $\mc^{-1}(E_k^U) =$
\[
 \bigcup_{\substack{\langle n_1, \dots, n_k \rangle 
  \in \N^k \\ n_i \neq n_j }} \left( \bigcap_{i = 1}^k \ev^{-1}_{n_i}U
\right) \cap\left(\bigcap_{n_i \neq n \in \N}
\ev^{-1}_{n}((0, 1]\setminus U)\right),
\]
a measurable set since it consists of countable unions and
intersections of generating elements.
\end{proof}

Given a discrete element-free distribution $\varphi$, we write
$\size{\varphi}$ for its size $\sum_{p \in \supp(\varphi)} \varphi(p)$, which
may be infinite. For example, if $\mu \in \G\N$, then $\size{\mc(\mu)}$ equals
the cardinality of the support of $\mu$.

\paragraph{Element-free distributions in ordered form.} It is common
in the probability literature to find element-free distributions formalized as decreasing
sequences over $[0, 1]$, rather than multisets (\cite{kingman-representation,pitman2006combinatorial}). We show that there is a map $\ord : \elemfreedisc \to \G\N$
which turns an element-free distribution into a distribution over $\N$
by enumerating the coefficients in decreasing order. This is possible
because the set $\supp(\varphi) \subseteq (0,1]$ under $\geq$ is
well-ordered: infinite increasing sequences of positive reals cannot sum to $1$. 

We formalize this to obtain a section (right-inverse) of multiplicity count. 
\begin{lemma}
  \label{lem:ord-disc-def}
  For $\varphi \in \elemfreedisc$, let $\ord(\varphi)$ be the distribution
  on $\N$ given by $\ord(\varphi)(n)  =$
  \[
    \begin{cases}
      \max \left\{ p \in \supp (\varphi) \mid \sum_{q \in
      \supp(\varphi), p \leq q} \varphi(q)> n \right\}
      & \text{if}\  \size{\varphi}> n \\
      0 & \text{otherwise.}
    \end{cases}
  \]
  Then $\ord$ is a measurable function $\elemfreedisc \to \G\N$ and a section of $\mc$.
\end{lemma}
\begin{example}
Let $\varphi$ be the multiset $[\frac{1}{5}, \frac{1}{5},
\frac{3}{5}]$ regarded as an element of $\elemfreedisc$. The
distribution $\ord(\varphi) \in \G\N$ is given by $0 \mapsto \frac{3}{5}$; $1, 2
\mapsto \frac{1}{5}$; and $n \mapsto 0$ for all $n \geq 3$.
\end{example}

\begin{proof}[Proof of Lemma~\ref{lem:ord-disc-def}]
It suffices to show that $\ord^{-1} V \in \Sigma_\elemfreedisc$ for every $V$
in any basis for the $\sigma$-algebra
$\Sigma_{\G\N}$. We consider the basis consisting of the subsets
$V^q_n  = \{ \mu \mid \mu \{n \}> q\}$, for $n \in \N$ and $q \in [0,
1]$.
(To see why this is a basis, note that if $\Sigma'$ is the
$\sigma$-algebra on $\G \N$ generated by the $V^q_n$, then each function $\ev_n :
\G \N \to [0, 1]$ is measurable, because the $(q, 1]$ generate $\Sigma_{[0,
  1]}.$ Since $\Sigma_{\G \N}$ is the smallest $\sigma$-algebra making
the $\ev_n$ measurable, and clearly $\Sigma' \subseteq \Sigma_{\G
  \N}$, they are equal.) The set $\ord^{-1}V^q_n$ consists of the $\varphi \in \elemfreedisc$
whose $n^\text{th}$ biggest element is above $q$, i.e.
\[
\ord^{-1}V^q_n = \bigcup_{k \geq n} E_k^{(q, 1]}
\]
which is measurable. So $\mc$ is measurable. 

The composite $\mc \circ \ord$ is the identity: for $\varphi \in
\elemfreedisc$ and $p \in (0, 1]$, we have that
$\varphi(p)$ is equal to the number of $n\in \N$ such that
\[
  p = \max \left\{ p' \in \supp(\varphi) \mid \textstyle\sum_{\substack{q \in \supp(\varphi)\\q\geq p'}} \varphi(q) > n \right\}.
\]
This number is precisely $\mc(\ord(\varphi))(p)$. 
\end{proof}

As a section, the map $\ord$ must be injective. It has measurable image: 
\begin{lemma}
  The image of $\ord :  \elemfreedisc \to \G\N$ is the subset 
  \[
 \{ \mu \in \G\N \mid \forall n \in \N. \  \mu\{n+1\} \leq \mu\{n\} \}
\]
of $\G\N$, which is in the $\sigma$-algebra $\Sigma_{\G\N}$.
\end{lemma}
\begin{proof}
 The subset above is the countable intersection $\bigcap_{i \in \N} \{ \mu \in \G\N \mid
  \mu(i + 1) \leq \mu(i) \}$, each component of which has measurable characteristic 
 function
    $\G\N \xrightarrow{\pairing{\ev_{i}}{\ev_{i+1}}}  \R \times \R
\xrightarrow{\geq} \{0, 1\}$.
\end{proof}

Measurable subsets of standard Borel spaces are standard Borel, and so
$\elemfreedisc$ is standard Borel.
As an aside, we note the following universal
characterization of $\elemfreedisc$ and $\mc$: 

\begin{proposition}
  The function $\mc$ is a coequalizer for the diagram consisting of
  all maps $\G(\alpha)$ for $\alpha \in \Aut(\N)$, the symmetry group
  of $\N$.
\[\begin{tikzcd}
	\G\N & \G\N && \elemfreedisc
	\arrow["{G(\alpha)}", shift left=2, from=1-1, to=1-2]
	\arrow["\cdots"{description}, draw=none, from=1-1, to=1-2]
	\arrow["\mc", from=1-2, to=1-4]
	\arrow[shift right=1, from=1-1, to=1-2]
	\arrow[shift right=2.5, from=1-1, to=1-2]
\end{tikzcd}\]
\end{proposition}
\begin{proof}
  The map $\mc$ forms a cone over the diagram, because counting
  multiplicity is invariant under any reindexing of $\N$. For the
  universal property, let $f : \G\N\to X$ be another invariant map
  into an arbitrary measurable space $X$. We must show that there
  exists a unique map $h : \elemfreedisc \to X$ such that $h\circ \mc =
  f$. Uniqueness follows from the fact that $\mc$, as a retraction, is
  epi. We set $h = f \circ \ord$, and it remains to show that
  \[
    \begin{tikzcd}
      \G\N & \elemfreedisc & \G\N \\
      & X
      \arrow["\mc", from=1-1, to=1-2]
      \arrow["\ord", from=1-2, to=1-3]
      \arrow["f"', from=1-1, to=2-2]
      \arrow["f", from=1-3, to=2-2]
    \end{tikzcd}
  \]
commutes. For every $\mu \in \G\N$, the distribution $\ord(\mc(\mu))$
is a reindexing of 
$\mu$ along some permutation $\alpha \in \Aut(\N)$,
i.e. $\mc(\ord(\mu))= \G(\alpha)(\mu)$. As $f = f \circ \G(\alpha)$ by the assumption
that $f$ is invariant, we are done.
  \end{proof}

So far we have considered discrete element-free distributions: infinite
multisets whose elements sum to $1$. These are a natural way to 
represent the element-free part of a probability distribution on
$\N$, or any other discrete space. 

\section{General element-free distributions}
\label{sec:general-element-free}

In this section we introduce the general notion of element-free
distribution, which also supports continuous distributions. 
We also develop two fundamental operations for
a standard Borel space $X$: multiplicity count for probability
measures on $X$ (\S\ref{sec:mult-count-gener}), and the sampling of new elements from a
base measure (\S\ref{sec:draw-elem-base-measure}). This gives a
retract situation (\S\ref{sec:space-nabla-retract}).

Recall that an element-free distribution is a multiset representing
the weights assigned to atoms of a probability
distribution. Distributions on continuous spaces can have a
nonatomic part, and so those weights may not sum to $1$. This
is the only difference between discrete and non-discrete element-free distributions.
\begin{definition}
  An \emph{element-free probability distribution} is a function $
    \varphi :
    (0, 1] \to \N $
  whose support is countable, and such that $\sum_{p \in \supp(\varphi)} \varphi(p)
  \cdot p \leq 1$. 
\end{definition}

We write $\elemfree$ for the space of element-free distributions, with
a basis of measurable subsets given by 
\[
D_k^U = \{ \varphi \in \elemfree \mid \sum_{\substack{p \in U \cap
    \supp(\varphi)}} \varphi(p) = k \} 
\]
for $U \in \Sigma_{(0, 1]}$ and $k \in \N$. (This extends the
$\sigma$-algebra of $\elemfreedisc$, in particular $E_k^U = D_k^U
\cap \elemfreedisc$.) 

Keeping the nonatomic part implicit, general element-free distributions can be
viewed as subprobability distributions on the natural numbers. 
It will sometimes be convenient to use generalized versions of the functions $\mc$ and $\ord$ from
Section~\ref{sec:discr-elem-free}. Recall that the space of sub-probability distributions
on $\N$ 
is written $\Gsub\N$, and contains $\G \N$ as a subspace. 
\begin{lemma}
 There are measurable functions $\mc : \Gsub \N \to \elemfree$ and
 $\ord : \elemfree \to \Gsub\N$ extending those given in
 Lemma~\ref{lem:mc-disc-def} and Lemma~\ref{lem:ord-disc-def}, and
 defined in the same way. Thus in particular $\elemfree$ is standard Borel.
\end{lemma}

We also note that $\elemfreedisc$ is a measurable subset of
$\elemfree$ : it is the pre-image of $\{ 1 \}$ under the measurable
function $
\elemfree \xrightarrow{\ord} \Gsub \N \xrightarrow{\ev_\N} [0, 1].$

\subsection{Multiplicity count for general spaces}
\label{sec:mult-count-gener}

By moving to general element-free distributions, we can consider
arbitrary probability measures on standard Borel spaces. We thus have
a new kind of multiplicity count, that forgets everything but the atom
weights (and their multiplicity).

For every standard Borel space $X$, this is given by
  \begin{align*}
    \mc : \G X &\longrightarrow \elemfree \\ 
   \mu &\longmapsto (r \mapsto \#\{ x \in X \mid \mu\{x\} = r\}).
  \end{align*}
\paragraph{Notational remark.} Here and throughout the paper we
overload the notation for our two fundamental operations $\mc$ and
$\base_\mu$ (next section), to emphasize that they correspond to the
same passage between element-free and ordinary probability. The types
should be clear from context, and so this should pose no confusion. 
 
\begin{example}
  \begin{itemize}
  \item For any Dirac $\delta_x \in \G X$, $\mc(\delta_x) = [1]$.
  \item If $\uniform$ is the uniform distribution on $[0, 1]$, then
    $\mc(\uniform)$ is the empty element-free distribution.
   \item We can have discrete-continuous mixtures. If $x \in [0, 1]$,
     then $\mc(\frac{1}{3} \delta_x + \frac{2}{3}\uniform) =
     [\frac{1}{3}]$, since there is a single atom with weight $\frac{1}{3}$.
  \end{itemize}
\end{example}

It is not immediate that $\mc$ is measurable. We first show a 
preliminary lemma:
\begin{lemma}
  \label{lem:epsilon}
  For $\varepsilon > 0$, define $\mc_\varepsilon : \G X \to \elemfree$
  by $\mc_\varepsilon(\mu)(r) = \mc(\mu)(r)$ if $r \geq \varepsilon$,
  and $0$ otherwise. Then $\mc_\varepsilon$ is measurable.
\end{lemma}

 The measurability of $\mc$ ought to follow because limits of
 measurable functions are measurable, but to make this limit argument
 precise one needs the topology of $\elemfree$. To avoid this, a direct argument is as follows: let $D^U_{\leq
    k} = \bigcup_{\ell \leq k} D^U_\ell$ and note that the $D^U_{\leq
    k}$ still generate $\Sigma_\elemfree$. We then have that 
  \begin{align*}
     &\mc^{-1}D^U_{\leq k} \\ &= \left\{ \mu \in \G X \mid \mu \text{ has $\leq
                            k$ atoms with weight in }U \right\}  \\
    &= \bigcap_{m \geq 1} \left\{ \mu \in \G X \mid \mu \text{ has $\leq k$
      atoms with weight in } U \cap \mathopen(\tfrac{1}{m}, 1 \mathclose]\right\} \\
     &= \bigcap_{m \geq 1} \mc_{\frac{1}{m}}^{-1} D^U_{\leq k},
  \end{align*}
     a countable union of measurable sets, and so $\mc$ is measurable.
\newcommand{\B}{\mathcal{B}}
\begin{proof}[Proof (of Lemma~\ref{lem:epsilon})]
  We only give details for a non-discrete $X$. Suppose w.l.o.g. that
  $X = [0, 1]$. There is an increasingly-refined sequence of partitions of $X$: for $m \in \N$ and
 $0 \leq n < 2^m$, let $B^n_{m} = [\frac{n}{2^m}, \frac{n+1}{2^m})$.

 We now show that, for every $\alpha \geq 0$ and $k \in \N$,
 the set $\mc^{-1}_\varepsilon D^{[\alpha, 1)}_k$ is
 measurable. We claim that this
 set is expressible using countable unions and intersections of basis
     elements, as follows:
     \[
       \bigcup_{\ell \in \N} \bigcap_{m \geq \ell}
       \bigcup_{\substack {\left\{n_1,...,n_k\!\right\} \\ 
           \subseteq \{0,\dots, 2m-1\!\}}}
      \left(
        \bigcap_{i=1}^{k}\ev^{-1}_{B^{n_i}_m}\left[\alpha,1\right] \right)
       \cap \left( \bigcap_{\substack{n \leq 2m-1\\ \forall i.\ n \neq n_i}}
         \ev^{-1}_{B^n_m}[0, \alpha) \right).
     \]
     To prove this claim  we use standard but verbose analytic
     arguments, which we omit. It remains to show that for every $U \in \Sigma_{(0, 1]}$ (and not just
for the $[\alpha, 1]$), the sets $\mc_{\varepsilon}^{-1} D^{U}_k$ are
measurable for $k \in \N$. Let $\mathcal{S}$ be
the set of those $U \subset [0, 1)$ satisfying this property. It suffices to show that
$\mathcal{S}$ is closed under relative complements and countable
increasing unions: by the $\pi$-$\lambda$ theorem (\eg~\cite[Thm.~1.8]{cinlar2011probability}) this implies
$\Sigma_{(0, 1]} \subseteq \mathcal{S}$, from which we can conclude.

For relative complements, observe that, whenever $U \subseteq V \in
\mathcal{S}$, $\mc_\varepsilon^{-1}D^{V \setminus U}_k = \bigcup_{m
  \in \N}\mc_\varepsilon^{-1}D^{V}_{k + m} \cap
\mc_\varepsilon^{-1}D^{U}_m$. This is a measurable set and so $V \setminus U
\in \mathcal{S}$. (Here we use that $\varepsilon > 0$, to avoid
infinite numbers of atoms.) For countable increasing unions, note that 
     $\mc_\varepsilon^{-1} D_k^{\bigcup_{n\in\N} U_n} = \bigcup_{m \in
       \N} \bigcap_{n \geq m} \mc_\varepsilon^{-1} D_k^{U_n}$, and so
     the proof is complete.
\end{proof}

In summary, the map $\mc : \G X \to \nabla$ is measurable for every
standard Borel space $X$. This is a deterministic way to forget
elements, and we now give a probabilistic way to recover them. 

\subsection{Drawing elements from a base measure}
\label{sec:draw-elem-base-measure}

We can turn an element-free distribution $\varphi \in \elemfree$ into an
ordinary distribution on a space $X$ by sampling elements from a base
measure $\mu \in \G X$. More precisely, the weights of
$\varphi$ are assigned to atoms sampled independently from $\mu$, and 
the nonatomic part in $\varphi$ is completed using $\mu$
itself. Formally, we will define a kernel $\base : \G X \times \elemfree \kerto \G X$. 

For each $\varphi \in \elemfree$, recall that the set $\supp(\varphi) \subseteq
[0,1]$ is countable and well-ordered under $\geq$, so we can enumerate its
elements in decreasing order: $\supp(\varphi) = \{ r_1\geq r_2 \geq \dots \}$.
Then, given any sequence $ \bm{x} = (x_i)_{i =
  1}^{\!\size{\varphi}}$ of elements in a set $X$, we can decompose 
the sequence $\bm{x}$ as
\[
\bm{x} = \bm{x}^{r_1} \oplus \bm{x}^{r_2} \oplus \dots  
\]
where $\oplus$ denotes concatenation, and each $\bm{x}^{r_i}$ is a subsequence of size
$\varphi(r_i)$. For each $r \in \supp(\varphi)$ we write the elements
of $\bm{x}^r$ as $(x^r_1, \dots, x^r_{\varphi(r)})$.
(Note that each operation $\bm{x} \mapsto \bm{x}^{r_i}$ is simply a
particular product projection, and thus a measurable function.)

\newcommand{\alloc}{\textsl{alloc}}
\begin{definition}[Allocation]
  For $\varphi \in \elemfree$ and $\bm{x} \in
    X^{\size{\varphi}}$, let
    \[ \alloc(\varphi, \bm{x}) = \sum_{r \in \supp(\varphi)} r\cdot \sum_{i =
        1}^{\varphi(r)} \delta_{x^r_i} \]
    be the subprobability measure that results from allocating
    elements in $\bm{x}$ to weights in $\varphi$. 
\end{definition}
\begin{definition}[Drawing elements from a base measure]
  \label{def:base-kernel}
  The kernel
    $\base_\mu : \elemfree \kerto \G X$ is defined for any $\varphi
    \in \elemfree$ and $V \in  \Sigma_{\G X}$ as follows:
  \begin{align*}
    \base_\mu(\varphi)(V) &= \\
    \int_{\bm{x} \in X^{\size{\varphi}}}
&\left[(1 - \total(\varphi)) \mu + \alloc(\varphi, \bm{x}) \in V \right]
\iid_{\size{\varphi}}(\mu)(\diff \bm{x}).
  \end{align*}
  (We denote by $[P]$ the indicator function of a property $P$.)
  \end{definition}

\begin{example} Let $\varphi$ be the multiset
  $[\frac{1}{3},\frac{2}{3}]$ viewed as an element of $\elemfree$. We give examples of
  sampling elements from discrete and continuous base measures.
  \begin{enumerate}
  \item \label{item1}
If $X = \{ 0, 1 \}$ and $\mu = \frac{1}{4} \delta_0 +
\frac{3}{4} \delta_1$, then the distribution
$\base_\mu(\varphi)$ on $\G X$ is purely atomic: it has four atoms
with weights given below.
\begin{align*}
\left( \frac{1}{3} \delta_0 + \frac{2}{3} \delta_1 \right)\
  \longmapsto \ \frac{3}{16} & & \delta_0
\  \longmapsto \ \frac{1}{16} \\
  \left( \frac{2}{3} \delta_0 + \frac{1}{3} \delta_1  \right)
  \  \longmapsto \ \frac{3}{16} && \delta_1
 \ \longmapsto \ \frac{9}{16}.
\end{align*}
\item If $X = [0, 1]$ and $\mu = \uniform$ is the uniform distribution, then $\base_\mu(\varphi)$ is given by
  \[
    \base_\mu(\varphi)(M) = \int_{(x, y) \in [0, 1]^2}
    \left[\left( \frac{1}{3} \delta_x + \frac{2}{3} \delta_y \right)\in M \right] \diff x
    \,\diff y
    \]
  for $M \in \Sigma_{\G  X}$.
\end{enumerate}
\end{example}

\begin{example}
  We now look at a different element-free
  distribution: $\psi = [\frac{3}{5}]  \in \elemfree$. Note that $\psi$ has a
  continuous part of weight $\frac{2}{5}$.
\begin{enumerate}
\item \label{item1}
If $X = \{ 0, 1 \}$ and $\mu = \frac{1}{4} \delta_0 +
\frac{3}{4} \delta_1$, then the distribution
$\base_\mu(\psi)$ on $\G X$ is purely atomic with atoms:
\begin{align*}
(\frac{2}{5} \mu + \frac{3}{5} \delta_0) \mapsto \frac{1}{4} && (\frac{2}{5} \mu + \frac{3}{5} \delta_1) \mapsto \frac{3}{4} 
\end{align*}
(These atoms reduce to 
$\frac{7}{10} \delta_0 + \frac{3}{10} \delta_1$ and $\frac{1}{10}
\delta_0  + \frac{9}{10} \delta_1$.)
\item If $X = [0, 1]$ and $\mu = \uniform$ is the uniform distribution, then
  \[
    \base_\mu(\psi)(M) = \int_{x\in [0, 1]}
    \left[\left( \frac{2}{5} \uniform + \frac{3}{5} \delta_x \right)\in M \right] \diff x
    \]
  for $M \in \Sigma_{\G  X}$.
\end{enumerate}
\end{example}

\subsection{The space $\nabla$ as a retract}
\label{sec:space-nabla-retract}

We can now move between the space of element-free distributions and 
spaces of distributions on  $X$ using the two constructions:
\[
  \begin{tikzcd}[column sep=4em]
  \G X \arrow{r}{\mc} & \elemfree &[-2em]  \elemfree
  \arrow[ker]{r}{\base_{\mu}} & \G X  \ \ \ (\mu \in \G X)
\end{tikzcd}    
 \]
 An important observation is that the composite
 \[
  \begin{tikzcd}[column sep=4em]
  \elemfree  
  \arrow[ker]{r}{\base_{\mu}} & \G X  \arrow{r}{\mc}  & \elemfree 
  \end{tikzcd}    
   \]
 is equal to the identity on $\elemfree$ whenever the measure $\mu$ is
 nonatomic. Intuitively, the elements produced by $\iid(\mu)$
 are all distinct with probability 1, and thus weights are never
 combined. (For a counterexample when $\mu$ is atomic: let $\mu =
 \delta_a$ on the set $\{ a, b \}$. Then for any $\varphi$, $\base_\mu(\varphi)$ assigns
 the element $a$ to every weight of $\varphi$, \ie~$\base_\mu(\varphi)
 = \delta_{\delta_a}$. Applying $\mc$ to $\delta_a$ returns $[1] \in
 \elemfree$. We see that the weights of $\varphi$ have all been combined.)

 \begin{proposition}
  \label{prop:uniform-base}
 For a standard Borel space $X$ and a nonatomic measure
 $\mu \in \G X$, the kernel $\base_\mu : \elemfree \kerto \G X$ is a
 section of $\mc : \G X \to \elemfree$ in the category of kernels; that is, $\mc \circ \base_\mu
 = \id_{\elemfree}$. 
\end{proposition}
\begin{proof}[Proof sketch]
The key idea is that, because $\mu$ is nonatomic, the measure $\iid(\mu) \in
 \G(X^\N)$ is concentrated on the subspace of sequences whose elements
 are pairwise distinct: let $S = \{ (x_n)_{n \in \N} \mid
 x_i \neq x_j \text{ if } i \neq j\}.$ For any $\bm{x} \in S$ (and, say, if
 $\varphi$ has infinite support) the
 distribution $\alloc(\bm{x}, \sigma)$ has the exact same atom weights
 as found in  $\varphi$.
\end{proof}

\section{Finite multisets and partitions}
\label{sec:multisets-partitions}

In this section we consider spaces of finite multisets and their
element-free counterpart: partitions of positive integers. Just like for distributions and
element-free distributions, we have two fundamental
operations: multiplicity count, a deterministic operation which turns a multiset of size $K$
into a partition of the number $K$ by forgetting elements; and  a probabilistic operation,
which turns a partition into a multiset by sampling elements from a base
measure. These operations will be defined in
\S\ref{sec:mult-count-base}. In \S\ref{sec:spac-finite-mult} we
recall the basics of finite multisets and partitions.

\subsection{Finite multisets and partitions}
\label{sec:spac-finite-mult}

\paragraph{Multisets.} We have previously used infinite multisets of positive reals to define
element-free distributions. Here we consider finite multisets over an
arbitrary space $X$.
\begin{definition}
  For a measurable space $X$, let $\M(X)$ be the set of finite
  multisets of elements of $X$. These are formally defined as
  functions $X \to \N$ with finite support. The $\sigma$-algebra $\Sigma_{\M(X)}$ is generated by the sets
  $
     E_k^U = \{ \varphi \in \M(X) \mid \sum_{U \cap \supp(\varphi)}
     \varphi(x) = k\} 
  $
  for $k \in \N$ and $U \in \Sigma_X$.
\end{definition}
\begin{notation}
For a multiset $\varphi \in \M(X)$, write $\size{\varphi}$ for its
size, \ie~$\size{\varphi} = \sum_{x \in \supp(\varphi)} \varphi(x)$.
For $K \in \N$, the subspace of multisets of size $K$ is denoted
$\M[K](X)$. (Note that $\M[K](X)$ is equal to the generating set $E^X_K$, and
thus a measurable subset of $\M(X)$.)
  \end{notation}
It it well-known (e.g.~\cite{DBLP:journals/corr/abs-2112-14048}) that if the space $X$ is standard Borel, then so is
$\M(X)$, and therefore so are all $\M[K](X)$.

\paragraph{Partitions} Partitions are defined as multisets of
integers. Each integer represents the size of a block of (unnamed)
elements. Partitions are a classical object of study in combinatorics
\cite{andrews1998theory}. The modern presentation we give now, in
terms of multisets, is due to Jacobs \cite{lics-jacobs}.

\begin{definition}
For a multiset $\sigma \in \M(\Ngz)$, the \emph{total} of $\sigma$ is
an integer $\total(\sigma)$ defined as $\total(\sigma) =
\sum_{n>0} \sigma(n) \cdot n$. 
\end{definition}

\begin{definition}
For $K \in \N$, the set of partitions
$\Part(K)$ is given by $\{\sigma \in \M(\Ngz) \mid \total(\sigma)= K \}.$
\end{definition}

\begin{example} The set of partitions $\Part(3)$ contains the three
  multisets $[1, 1, 1]$, $[1, 2]$, and $[3]$, representing the possible combinations of block sizes in
    partitions of a  three-element set $\{ a, b, c\}$: 
    $\{\{a\}, \{b\}, \{c\}\}$, $\{\{a, b\}, \{c\}\}$,
$\{\{a\}, \{b, c\}\}$,
$\{\{a, c\}, \{b\}\}$,
$\{\{a, b, c\}\}$.
\end{example}

The sets $\Part(K)$ are all finite, and we regard them as discrete measurable spaces.

Before proceeding, we recall some combinatorial coefficients
for multisets and partitions, taken from \cite{lics-jacobs}. 
\begin{definition}
For a multiset $\varphi \in \M(X)$ over any set $X$, 
  \begin{itemize}
\item The multinomial coefficient $\multicoeff{\varphi}$ is defined as
  $\frac{\size{\varphi}!}{\prod_{x \in \supp(\varphi)}\varphi(x)! }$.
 \item For $n \geq
   \size{\varphi}$, the coefficient ${n \choose \varphi}$ is defined as
   \[
       {n \choose \varphi} =  \multicoeff{\varphi} \cdot {n \choose \size{\varphi}} =
        \frac{n!}{(n - \size{\varphi})! \cdot \prod_{x} \varphi(x)!}
     \]
 \end{itemize}
Finally, for a partition $\sigma \in \Part(K)$, the coefficient
 $\multicoeffpart{\sigma}$ is defined as $\frac{K!}{\prod_{n \leq K}(n!)^{\sigma(n)}}$.
\end{definition}

\subsection{Multiplicity count for multisets}
\label{sec:mult-count-base}
Every finite multiset over a space $X$ induces a partition, obtained
by forgetting the elements. For example, for $X = \{ a, b, c\}$, 
the multiset $[a, a, b]$ gives the partition $[2, 1]$.

In other words, partitions are
element-free multisets. This operation is another form of
multiplicity count \cite{lics-jacobs}:
\begin{align*}
  \mc : \M[K](X) &\longrightarrow \Part(K) \\ 
  \varphi \ \  &\longmapsto \  \left(n \mapsto  \#\{ x \in \supp(\varphi) \mid \varphi(x) = n \}\right)
\end{align*}

For our applications we note that multiplicity count over standard
Borel spaces is measurable. 
\begin{lemma}
  For every standard Borel space $X$ and integer $K \geq 0$, the
  function
  $\mc : \M[K](X) \to \Part(K)$ is measurable.  
\end{lemma}
\begin{proof}[Proof] We can leverage the fact that $\mc : \G X
  \to \elemfree$ is measurable. The idea is to regard $\M[K](X)$
  as a subspace of $\G X$, via the injective map $f_k: \M[K](X) \to \G
  X$ given by $f_K(\varphi) = \sum_{x \in
    \supp(\varphi)}\frac{\varphi(x)}{K} \delta_x$.
  For $U \in \Sigma_X$, the following diagram commutes
  \[
    \begin{tikzcd}
      \M[K](X) \arrow[swap]{d}{\ev_{U}} \arrow{r}{f_K} & \G X   \arrow{d}{\ev_U} \\
      \N \arrow{r}{\frac{\cdot}{K}} & {[0, 1]}
    \end{tikzcd}
  \]
  and so each $\ev_U \circ f_K$ is measurable. Since the $\ev_U$ generate
  $\Sigma_{\G X}$, $f_K$ is measurable. Similarly, via the map $\Part(K) \to \elemfree$ which divides every
  element of a partition by $K$, we
   identify the set $\Part(K)$ with the subset of $\elemfree$
  consisting of multisets whose support is contained in $\{
  \frac{i}{K} \mid 1 \leq i \leq K \}$. This is a finite and thus
  measurable subset. The diagram below commutes
  \[
    \begin{tikzcd}
      \M[K](X) \arrow[swap]{d}{f_K} \arrow{r}{\mc} & \Part(K) \arrow{d}{f} \\
      \G X \arrow{r}{\mc} & \elemfree
    \end{tikzcd}
  \]
  and so the lemma holds.
  \end{proof}

\subsection{Partitions and base measures}
  
In the reverse direction, given a partition, we can randomly sample an
element for each block to reconstruct a multiset. We give a formal
definition, and then a more combinatorial characterization. For this
section we fix a standard Borel space $X$, a probability measure
$\mu\in \G X$, and an integer $K \geq 0$.

We first make an observation similar to that in \S\ref{sec:draw-elem-base-measure}: for a partition $\sigma\in \Part(K)$, any 
$\bm{x} \in X^{\size{\sigma}}$ is necessarily of the form 
\[
\bm{x} = \bm{x}^1 \oplus \dots \oplus \bm{x}^K
  \]
for (possibly empty) subsequences $\bm{x}^n = (x^n_1, \dots,
x^n_{\sigma(n)})$. (In other words, we consider the blocks of $\sigma$
in increasing order of size.)
\begin{definition}[Allocation]
  For $\sigma \in \Part(K)$ and $\bm{x} \in X^{\size{\sigma}}$, define
  \[
    \alloc(\sigma, \bm{x}) = \sum_{n =1}^K \sum_{i = 1}^{\sigma(n)}
    n [x^n_i] \quad \in \M[K](X)
  \]
  where the sum operation and $\N$-action on multisets are defined pointwise on
  the corresponding functions $X \to \N$. (So, in particular, $n  [x] = [x, \dots,
  x]$ with $x$ occurring $n$ times.)
\end{definition}
\begin{definition}[Base measure draws for partitions]
  For $\sigma\in \Part(K)$, the measure
  $\base_\mu(\sigma)$ on $\M[K](X)$ is defined by 
  \[
\base_\mu(\sigma)(V) = \int_{\bm{x} \in X^{\size{\sigma}}} \left[
  \alloc(\sigma, \bm{x})\in V \right] \iid_{\size{\sigma}}(\mu)(\diff \bm{x}).
\]
 for $V \in \Sigma_{\M[K](X)}$. This defines a kernel $\base_\mu :  \Part(K) \kerto \M[K](X)$. 
\end{definition}

We have another retract situation between multisets over a continuous
space and partitions. 
\begin{lemma}
  \label{lem:multi-part-retract}
For a non-atomic $\mu \in \G X$, and $K \geq 0$, the kernels $\base_\mu :
\Part(K) \kerto \M[K](X)$ and $\mc : \M[K](X) \kerto \Part(K)$ form
a section-retraction pair, \ie~$\mc \circ \base_\mu = \id_{\Part(K)}$.
\end{lemma}

It will be helpful to have a more
combinatorial expression of the kernel $\base_\mu$. We recall the following result.
\begin{lemma}[Dash and Staton \cite{DBLP:journals/corr/abs-2101-10479}]
  \label{lem:ring}
  The family of subsets of $\M(X)$ of the form 
  \[
     \bigcup_{i \in I} \bigcap_{j \in J} E^{U_{j}}_{k_{i, j}} 
   \]
   where $I$ and $J$ are countable, the $U_j$ are all disjoint
   measurable subsets of $X$, each $k_{i, j} \in \N$, and the
   sequences $(k_{i, j})_{j \in J}$ are distinct, forms a \emph{ring} which generates $\Sigma_{\M(X)}$.
 \end{lemma}
 
A measure on any measurable space is determined by its values on
a generating ring \cite{cinlar2011probability}, and so the following lemma suffices to characterize
$\base_\mu(\sigma)$ for any $\sigma \in \Part(K)$.
\begin{lemma}
For $\sigma \in \Part(K)$, 
\begin{align*}
& \base_\mu(\sigma)(\bigcup_{i \in I} \bigcap_{j \in J} E_{k_{i,
  j}}^{U_j})  =
\sum_{i \in I} \\ 
& \sum_{\substack{(\tau_j)_{j \in J}, \sum_j \tau_j  \leq \sigma \\
    \total(\tau_j) = k_{i, j}}}  
  \mu(V)^{\size{\sigma}-\size{\sum_j \tau_j}}
  \left(\prod_{j \in J} \mu(U_j)^{\size{\tau_j}}\right)
  \prod_{n \leq K}  {\sigma(n )\choose (\tau_j(n))_{j\in J}}
\end{align*}
where $V$ is the complement of the union $\bigcap_{j \in J} U_j$, and where the order $\leq$ on partitions is defined pointwise on the
corresponding functions $\Ngz \to \N$. 
\end{lemma}

In this section we have defined the two fundamental operations $\mc : \M[K](X) \to \Part(K)$
and $\base_\mu : \Part(K) \kerto \M[K](X)$ for moving between the
element-based and element-free settings.

\section{Sampling multisets and partitions}
\label{sec:sampl-mult-part}
In this paper, the process of generating finite collections of
independent and identically distributed samples is modelled by the
multinomial kernels $\mn_K : \G X \kerto \M[K](X)$, for $K \in
\N$. This is the usual kind of sampling, with elements. In this
section we contrast it with an element-free version, defined by the
partition multinomial kernels $\pmn_K : \elemfree \kerto \Part(K)$. As we will see, the
diagrams
\begin{equation}
  \label{eq:3}
    \begin{tikzcd}
	{\G X} & \elemfree \\
	{\M[K](X)} & {\Part(K)}
	\arrow["{\mn_K}"', ker, from=1-1, to=2-1]
	\arrow["\mc", from=1-1, to=1-2]
	\arrow["\mc"', from=2-1, to=2-2]
	\arrow["{\pmn_K}", ker, from=1-2, to=2-2]
      \end{tikzcd}\quad \ 
     \begin{tikzcd}[column sep=3em]
	{\G X} & \elemfree \\
	{\M[K](X)} & {\Part(K)}
	\arrow["{\mn_K}"', ker, from=1-1, to=2-1]
	\arrow["\base_\mu"', ker, from=1-2, to=1-1]
	\arrow["\base_\mu", ker, from=2-2, to=2-1]
	\arrow["{\pmn_K}", ker, from=1-2, to=2-2]
      \end{tikzcd}
\end{equation}
both commute in $\Kl(\G)$.

\subsection{Sampling with elements}
\label{sec:sampl-with-elem}

We first discuss the generation of multisets of iid samples. The function
$\iid_K : \G X \to \G (X^K)$ sending $\mu$ to the $K$-ary product measure
$\mu^K$ is regarded as a kernel $\G X\kerto X^K$ that generates ordered lists of iid samples. We are interested in
the composite \[
  \mn_K : \G X \kerto X^K \to M[K](X),\] where the second
(deterministic) map forgets the list order. This composite is
characterized as follows: 
\begin{lemma}
Let $\mu \in \G X$, for a standard Borel space $X$. For the generating
sets of Lemma~\ref{lem:ring}, if $\sum_{j} k_{i, j} \leq K$
for all $i\in I$,  then
\begin{align*}
\mn_K(\mu) &\left( \bigcup_{i \in I} \bigcap_{j \in J}
  E^{U_{j}}_{k_{i, j}} \right)  
  \\
  &= \sum_{i \in I} \frac{K!}{(K - \sum_{j}
  k_{i, j})! \prod_j k_{i, j}!} \mu(V)^{K -
  \sum_{j} k_{i, j}} \prod_j \mu(U_j)^{k_{i, j}}.
\end{align*}
\end{lemma}

\subsection{Sampling without elements}
\label{sec:sampl-without-elem}

We now turn to the generation of iid random partitions directly from
an element-free distribution. This process was studied in depth by
Jacobs \cite{lics-jacobs} in the restricted setting of finitely-supported, discrete element-free 
distributions. Our results rely and expand on his `partition multinomial' sampling formula, which we first recall:

\begin{definition}[{\cite[Def.~15]{lics-jacobs}}]
  \label{def:iid-disc}
Let $\varphi$ be a discrete element-free distribution with finite
support $\{ r_1, \dots, r_\ell\}$, and with $\varphi(r_i) = n_i$. The
distribution $\pmn_K(\varphi)$ on $\Part(K)$ is given by 
\[
 \sigma \mapsto \multicoeffpart{\sigma} \sum_{\substack{\sigma_1, \dots,
     \sigma_\ell \\ \size{\sigma_i} \leq 
     n_i, \sum_i \sigma_i = \sigma }} \prod_{i} {n_i\choose \sigma_i}\cdot r_i^{\total(\sigma_i)}.
\]
\end{definition}
The intuitive justification for this formula is that we must consider
all the possible ways to allocate blocks of $\sigma$ to coefficients
in $\varphi$. We first look at all possible ways of splitting $\sigma$
into sub-partitions $\sigma_{i}$. The blocks of each $\sigma_i$ are
then allocated to distinct copies of the coefficient $r_i$ in
$\varphi$; there are ${n_i \choose \sigma_i}$ possible such
allocations. 

We now proceed
to generalize this formula to the full space of element-free
distributions. The first step is to account for infinitely 
supported (but still discrete) element-free distributions,  which we can do by rewriting the above as follows:
\begin{equation}
  \label{eq:5}
 \sigma \mapsto \multicoeffpart{\sigma}
\sum_{\substack{(\sigma_r)_{r \in \supp(\varphi)}  \\\size{\sigma_r} \leq
     \varphi(r), \sum_r \sigma_r = \sigma}} \prod_{r \in \supp(\varphi)} {\varphi(r) \choose \sigma_r}\cdot r^{\total(\sigma_r)}
\end{equation}
where the index of the sum is countable, even if $\supp(\varphi)$ is infinite, because all but finitely many
$\sigma_r$ must be empty by the condition that $\sum_r \sigma_r = \sigma$. (Similarly, in the infinite product all but
finitely many factors are equal to $1$.)

We now generalize this formula to general element-free distributions, whose
coefficients may not sum to 1. Recall that the continuous part of
$\varphi$ represents an atomless distribution, and therefore only
produces singleton blocks. In the definition below, the idea is
to sum over the possible numbers $k$ of singleton blocks in $\sigma$
arising from this continuous part. 

\begin{definition}
  Let $\varphi \in \elemfree$, and let $w \in [0, 1]$ be the
  weight of its continuous part, i.e.~$w= 1- \sum_{r \in \supp(\varphi)} r\cdot \varphi(r)$.

  For $K \geq 0$, the distribution $\pmn_K(\varphi)$ over $\Part(K)$ is given by
\[
 \sigma \mapsto  \multicoeffpart{\sigma}
  \sum_{k = 0}^K  \frac{w^k}{k!} \sum_{\substack{(\sigma_r)_{r \in \supp(\varphi)} \\ \size{\sigma_r} \leq
    \varphi(r)\\ \sum_{r} \sigma_r + k [1] = \sigma}}
 \prod_{r \in \supp(\varphi)} {\varphi(r)\choose \sigma_r}\cdot r^{\total(\sigma_r)}
\]
for $\sigma \in \Part(K)$, where $k [1] \in \Part(k)$ is the
all-singleton partition.
\end{definition}
It is easy to see that if $w = 0$ the formula reduces to the
expression in \eqref{eq:5}. We must now show that $\pmn_K(\varphi)$
is a well-defined probability distribution.

We apply a similar proof method as Jacobs, based on 
a `partitions multinomial theorem'
(\cite[Theorem~16]{lics-jacobs}). We first extend the theorem to an infinite
 setting using a straightforward limit argument.  
 \begin{lemma}[Infinite partitions multinomial]
   \label{lem:part-multinomial}
  Let $I$ be a countable set, $(r_i)_{i \in I}$ a collection
  of non-negative reals, and  $(n_i)_{i\in I}$ a collection of non-negative integers. For $K \geq 0$, 
  \begin{align*}
    \left(\sum_{i\in I} r_i \cdot n_i \right)^K = \sum_{\sigma
    \in \Part(K)}  \multicoeffpart{\sigma}
\sum_{\substack{(\sigma_i)_{i \in I}  \\\size{\sigma_i} \leq
     n_i \\\sum_{i \in I} \sigma_i = \sigma}} \prod_{i \in I} {n_i \choose \sigma_i}\cdot r_i^{\total(\sigma_i)}.
  \end{align*}
\end{lemma}
\begin{proof}
Countable sums of positive reals are obtained as supremums of finite
sums, and we can apply the finite partitions multinomial theorem from
\cite{lics-jacobs}, $ \left(\sum_{i\in I} r_i \cdot n_i \right)^K $
\begin{align*}
 &\overset{(1)}=  \sup_{J \subseteq_f I} \left(\sum_{j \in J} r_j \cdot n_j\right)^K \\
  &\overset{(2)}=  \sup_{J \subseteq_f I} \left(\sum_{\sigma \in \Part(K)}
    \multicoeffpart{\sigma} \sum_{\substack{(\sigma_j)_{j \in J}\\
  \size{\sigma_j} \leq n_j, \sum_{j} \sigma_j = \sigma   }}
  \prod_{j \in J} {n_j \choose \sigma_j} \cdot r_j^{\total(\sigma_j)}
  \right) \\
    &\overset{(3)}=  \sum_{\sigma \in \Part(K)}
    \multicoeffpart{\sigma} \sup_{J \subseteq_f I} \left(\sum_{\substack{(\sigma_j)_{j \in J}\\
  \size{\sigma_j} \leq n_j, \sum_{j} \sigma_j = \sigma   }}
  \prod_{j \in J} {n_j \choose \sigma_j} \cdot r_j^{\total(\sigma_j)}
  \right) \\
    &\overset{(4)}=  \sum_{\sigma \in \Part(K)}
    \multicoeffpart{\sigma} \sum_{\substack{(\sigma_i)_{i\in I}\\
  \size{\sigma_i} \leq n_i, \sum_{i} \sigma_i = \sigma   }}
  \prod_{i \in I} {n_i \choose \sigma_i} \cdot r_i^{\total(\sigma_i)},
\end{align*}
where $(1)$ is because taking the $K$th power is continuous and
increasing, $(2)$ is by the finite partitions multinomial theorem,
$(3)$ is because sums and sups can be interchanged over the positive
reals, and $(4)$ uses the fact that any $(\sigma_i)_{i \in I}$ as in
the sum index is necessarily $0$ outside of a finite $J \subseteq I$.
\end{proof}

\begin{proposition}
For $\varphi \in \elemfree$ and $K \geq 0$, $\pmn_K(\varphi)$ is a
well-defined probability distribution.
\end{proposition}
\begin{proof}
We first observe that for $k \leq K$ and $\tau \in \Part(K - k)$,
  \[
    {K \choose k} \multicoeffpart{\tau} = \frac{1}{k!}
    \multicoeff{\tau + k[1]}.
  \]
(This is easy to verify directly.) We then write $w$ for the weight of the
continuous part of $\varphi$. We apply the binomial theorem,
 Lemma~\ref{lem:part-multinomial}, and the identity above:  
  \begin{align*}
1 &=  \left(w + \sum_{r \in \supp(\varphi)} r \cdot \varphi(r)
    \right)^K \\
    &=   \sum_{k = 0}^K {K \choose k} w^k \left( \sum_{r \in \supp(\varphi)} r
    \cdot \varphi(r) \right)^{K-k} \\
  &=     \sum_{k = 0}^K {K \choose k} w^k
    \sum_{\tau \in \Part(K-k)} \multicoeffpart{\tau} \sum_{(\tau_r)_{r
    \in \supp(\varphi)}}
    \prod_{r} {\varphi(r) \choose \tau_r} r^{\total(\tau_r)} \\
  &=     \sum_{k = 0}^K \frac{w^k}{k!}
    \sum_{\tau \in \Part(K-k)} \multicoeffpart{\tau + k[1]} \sum_{\substack{(\tau_r)_{r
    \in \supp(\varphi)} \\ \size{\tau_r} \leq \varphi(r), \sum_r{\tau_r} = \tau }}
    \prod_{r} {\varphi(r) \choose \tau_r} r^{\total(\tau_r)} \\
    &= \sum_{\sigma \in \Part(\sigma)} \multicoeffpart{\sigma}
      \sum_{k = 0}^K \frac{w^k}{k!}
      \sum_{\substack{(\sigma_r)_{r     \in \supp(\varphi)} \\ \sigma_r
    \leq \varphi(r)\\ \sum_r \sigma_r + k[1] = \sigma }}
    \prod_{r} {\varphi(r) \choose \sigma_r} r^{\total(\sigma_r)}
  \end{align*}
  and we have recovered the sum of coefficients in $\pmn_K(\varphi)$.
\end{proof}

\subsection{Relating the two forms of sampling}

We can now make formal the relationship between ordinary sampling and
element-free sampling. The first key result is that multiplicity count
commutes with multinomial distributions. 

\begin{restatable}{proposition}{mccommuteiid}
\label{prop:mc-commute-iid}
  Multiplicity count commutes with multinomial sampling, i.e.~the diagram on the
left of equation \eqref{eq:3} commutes.
\end{restatable}
\begin{proof}
The crux of the argument is the same as in the 
the restricted setting of Jacobs
\cite[Prop.~17~(2)]{lics-jacobs}.
\end{proof}

The second key result is that $\base_\mu$ also commutes with
multinomial distributions. Here we give only the discrete proof, which
contains the main combinatorial argument. We make use of the
well-known multinomial formula for computing finite powers of sums:
\[ \left(\sum_{i \in I} a_i \right)^N = \sum_{\chi \in \M[N](I)} \multicoeff{\chi}
  \prod_{i} a_i^{\chi(i)}. \]

\begin{restatable}{proposition}{basecommuteiid}
  \label{prop:base-commute-iid}
  Sampling elements from a base measure $\mu$ commutes with multinomial sampling, i.e.~the diagram on the
right of equation \eqref{eq:3} commutes.
\end{restatable}
\begin{proof}[Proof (simple case)]
 Fix $\varphi \in \elemfreedisc$. Consider a family of disjoint sets
 $U_j  \in \Sigma_X$ and a family $k_{j} \leq K$, for $j \in J$. %
   Assume w.l.o.g. that $\sum_j k_{j} = K$, and consider the set
   $\bigcap_j E^{U_j}_{k_{j}}$. By definition,
  \begin{align*}
 &   (\mn_K \circ
    \base_\mu)(\varphi)\left(\bigcap_j E^{U_j}_{k_{j}}\right) \\
    &= \int_{\bm{x} \in X^{\size{\varphi}}}
      \mn_K\left(\sum_{\substack{r \in \supp(\varphi)\\ 1
      \leq h \leq \varphi(r)}} r \cdot \delta_{x^r_h}\right)\left(\bigcap_j E^{U_j}_{k_{j}}\right) 
      \iid_{\size{\varphi}}(\mu)(\diff \bm{x}) \\
 &= {K \choose (k_{j})_{j \in J}}\int_{\bm{x} \in X^{\size{\varphi}}} \prod_{j \in J} \left( \sum_{r, 1
      \leq h \leq \varphi(r)} r \cdot \delta_{x^r_h}(U_j) \right) ^{k_{j}}
   \iid_{\size{\varphi}}(\mu)(\diff \bm{x}).
  \end{align*}
By the multinomial formula, this is equal to 
  \begin{align*}
    {K \choose (k_{j})_{j}}& \sum_{\substack{(\chi_j)_j, \ \chi_j \in \M[k_{j}]\\( \{ (r, h) \mid h \leq
      \varphi (r)\})}}  \left(\prod_j \multicoeff{\chi_j}\right)
 \\ &  \int_{\bm{x} \in X^{\size{\varphi}}}  \prod_{\substack{j\in J,r,
    \\ 1\leq
    h\leq \varphi(r)}} (r \cdot \delta_{x^r_h}(U_j))^{\chi_j(r, h)}
      \iid_{\size{\varphi}}(\mu)(\diff \bm{x})  
  \end{align*}
which we simplify to 
  \begin{align*}
   {K \choose (k_j)_{j}} \sum_{\substack{(\chi_j)_j, \ \chi_j \in \M[k_{j}]\\( \{ (r, h) \mid h \leq
    \varphi (r)\}) \\ (\chi_j) \text{\ disjoint}}} \bigl(\prod_j
    \multicoeff{\chi_j}
    &\mu(U_j)^{\left|\supp\left(\chi_j\right)\right|}\bigr) \\[-3em]
  & \prod_{r} r^{\sum_{j, 1\leq h \leq \varphi(r)}\chi_j(r, h)} 
      \iid_{\size{\varphi}}(\mu)(\diff \bm{x})
  \end{align*}
by inspecting the integral, noting in particular that it is 0 when $(\chi_j)$ have overlapping support.

  Now observe that we can turn any family $(\chi_j)_j$ (as in the
  sum indices above) into a tuple $((\sigma_r)_{r \in \supp(\varphi)},
  (\tau_j)_{j \in J})$. Formally each $\sigma_r$ is $\mc$ applied to the multiset
  $\sum_j \chi_j$ restricted to pairs of the form $(r, h)$, and $\tau_j$ 
  is simply defined as $\mc(\chi_j)$. So in particular $\sum_{r} \sigma_r
  \in \Part(K)$, $\tau_j \in \Part(k_j)$, $\size{\sigma_r} \leq \varphi(r)$, and $\sum_j \tau_j = \sum_{r} \sigma_r$.

 In this change-of-variable operation, we note that
  \[{K \choose (k_{j})_{j \in J}} \prod_{j \in J} \multicoeff{\chi_j}=
  \multicoeffpart{\textstyle \sum_{r} \sigma_r},\] and that each pair
  $((\sigma_r)_r, (\tau_j)_j)$ satisfying the conditions of the previous
  sentence has precisely $\prod_{n \leq K} {\sigma(n) \choose \tau_j(n), j\in J} \prod_{r}
  {\varphi(r) \choose \sigma_r}$ antecedents. We note also that
  $\left|\supp(\chi_j)\right| =  \size{\tau_j}$.
  Therefore we can rewrite the above expression as:
  \begin{align*}
    \begin{split}
       \sum_{\substack{(\sigma_r)_{r \in
      \supp(\varphi)}\\
      \size{\sigma_r} \leq \varphi(r)}}
{    \multicoeffpart{ \textstyle \sum_r \sigma_r}}
    \sum_{\substack{(\tau_j)_{j} \\ \sum_j \tau_j = \sum_r \sigma_r \\ \total(\tau_j) = k_{j}}}
    \biggl(\prod_j &\mu(U_j)^{\size{\tau_j}} \prod_r
    r^{\total(\sigma_r)}\biggr) \\[-2em]
    & \prod_{n \leq K} {\sigma(n) \choose (\tau_j(n))_{j\in J}} \prod_{r}
  {\varphi(r) \choose \sigma_r} \\
    \end{split} \\
    \begin{split}
    = \sum_{\sigma \in \Part(K)}
    \sum_{\substack{(\sigma_r)_{r \in
  \supp(\varphi)}\\
    \sum_r \sigma_r = \sigma \\
    \size{\sigma_r} \leq \varphi(r)}}
    \multicoeffpart{ \sigma}
    \sum_{\substack{\tau_j \\ \sum_j \tau_j = \sigma \\ \total(\tau_j) = k_{j}}}
     \biggl(&\prod_j \mu(U_j)^{\size{\tau_j}} \prod_r
      r^{\total{\sigma_r}}\biggr) \\[-2em]
     & \prod_{n \leq K} {\sigma(n) \choose (\tau_j(n))_{j\in J}} \prod_{r}
      {\varphi(r) \choose \sigma_r} \\
    \end{split} \\
\begin{split}
=  \sum_{\sigma \in \Part(K)}   \multicoeffpart{\sigma}
 & \biggl(\sum_{\substack{(\sigma_r)_{r \in
      \supp(\varphi)}\\    \sum_{r} \sigma_r = \sigma \\
  \size{\sigma_r} \leq \varphi(r)}}
  \prod_r {\varphi(r) \choose
    \sigma_r} r^{\total{\sigma_r}} \biggr)  \\
   & \biggl(
     \sum_{\substack{\tau_j \\ \sum \tau_j = \sigma \\ \total(\tau_j) = k_{j}}}
    \prod_j \mu(U_j)^{\size{\tau_j}} \prod_{n \leq K} {\sigma(n) \choose (\tau_j(n))_{j\in J}}
  \biggr)  \\
\end{split}\\
&  =  \sum_{\sigma \in \Part(K)} \pmn_K(\varphi)(\sigma)
    \cdot \base_\mu(\sigma)\left(\bigcap_j E^{U_j}_{k_{j}}\right)  \\
 &  = (\base_\mu \circ \pmn_K)(\varphi)\left(\bigcap_j E^{U_j}_{k_{j}}\right) 
  \end{align*}
  and so the argument is complete.
 \end{proof}
 
These two propositions make precise the informal commutative diagrams in
\S\ref{sec:rand-part-bayes} and \S~\ref{sec:base-measures-random}, and
together clarify the relationship between ordinary element-based
sampling, and element-free sampling.

\section{The representation of infinite random partitions}
\label{sec:kingman}

In this section we turn to Kingman's representation theorem for random
partitions. This theorem is concerned with infinite random partitions,
\ie~random partitions of $\N$, formalized as infinite sequences of
random \emph{finite} partitions subject to a consistency
condition, which states that deleting a random point from the
random partition at level $K+1$ should yield the random partition at level $K$.

\subsection{Draw-delete maps for multisets and partitions}
\label{sec:draw-delete-maps}

In order to state the theorem with the appropriate consistency
condition, we recall the definitions of \emph{draw-delete} kernels,
both with and without elements (respectively, for multisets and for
partitions). These represent the process of deleting an element from a
multiset in $M[K+1](X)$, or a point from a partition in $\Part(K+1)$,
uniformly chosen among all $K+1$. 

We first fix some notation. For $\varphi \in \M[K + 1](X)$, and
$x \in \supp(\varphi)$, we write $\varphi^{-x} \in \M[K](X)$ for the
multiset obtained from $\varphi$ by removing a copy of $x$. Formally
$
\varphi^{-x}(x') = \varphi(x') - [x = x'], 
$
for $x' \in X$. 
Similarly, for $\sigma \in \Part(K+1)$, and $n \in \supp(\sigma)$, the
partition $\sigma^{n-} \in \Part(K)$ is obtained by removing an
element from a block of size $n$ in $\sigma$. Formally, for $n' > 0$, 
\[
  \sigma^{n-}(n') = \begin{cases}
                      \sigma(n') -1 & \text{ if }n' = n \\ 
                      \sigma(n') + 1 &\text{ if }n' = n - 1 \\
                      \sigma(n') & \text{otherwise.}
                    \end{cases}
                  \]

                  \begin{lemma}
                    \label{lem:dd-commute-mc}
  Let $X$ be a standard Borel space, and $K \geq 0$. There are kernels 
     $\dd : \M[K + 1](X) \kerto \M[K](X)$ and $\pdd : \Part(K + 1)
    \kerto \Part(K)$ given by 
\[
    \dd(\varphi) =  \textstyle \sum_{x \in \supp(\varphi)}
       \frac{\varphi(x)}{\size{\varphi}} \cdot \delta_{\varphi^{-x}}
       \quad 
              \pdd( \sigma)
=
   \textstyle   \sum_{n \in \supp(\sigma)} \frac{n \cdot
     \sigma(n)}{\total(\sigma)}\cdot \delta_{\sigma^{n-}}
 \]
 for $\varphi \in \M[K + 1](X)$ and $\sigma \in \Part(K + 1)$.

 Moreover the
 following diagram commutes in $\Kl(\G)$. 
 \[
   \begin{tikzcd}
      \M[K + 1](X) \arrow[ker]{r}{\dd} \arrow[swap]{d}{\mc} & \M[K](X) \arrow{d}{\mc}\\
        \Part(K + 1) \arrow[ker, swap]{r}{\pdd} & \Part(K)
     \end{tikzcd}
   \]
  \end{lemma}
\begin{proof}
This is \cite[Lemma 7]{lics-jacobs}. We must additionally check the measurability
of $\dd : \M[K+1](X) \kerto \M[K](X)$. For this we note that $\dd$ is
equal to a composite 
$\M[K + 1](X) \kerto X^{K + 1} \kerto X^K \to \M[K](X)$
where the first kernel picks a random enumeration, the
second one performs draw-and-delete on sequences, and the third map
turns the resulting sequence back into a multiset.  
\end{proof}  

We now show the corresponding property for $\base_\mu$.
\begin{lemma}
For every $K$, the following diagram commutes in $\Kl(\G)$.
  \[
   \begin{tikzcd}
        \Part(K + 1)  \arrow[ker,swap]{d}{\base_\mu}  \arrow[ker]{r}{\pdd}
        & \Part(K)  \arrow[ker]{d}{\base_\mu} \\
     \M[K + 1](X) \arrow[ker, swap]{r}{\dd}& \M[K](X) 
     \end{tikzcd}
   \]
\end{lemma}
\begin{proof}
Let $\sigma \in \Part(K + 1)$ and let $\bm{x} \in
X^{\size{\sigma}}$. Let 
$n_1 \leq \dots \leq n_{\size{\sigma}}$ be the (unique) increasing enumeration of
the blocks of $\sigma$. A straightforward calculation gives
\[
  \dd(\alloc(\bm{x}, \sigma)) = \sum_{i = 1}^{\size{\sigma}} \frac{n_i}{K + 1}
  \delta_{\alloc(\bm{x}, \sigma)^{-{x_i}}},
\] and thus for $V \in
\Sigma_{\M[K](X)}$, $(\dd\circ \base_\mu)(V) = $
\begin{align*}
&  \sum_{i = 1}^{\size{\sigma}}  \frac{n_i}{K + 1}
                          \int_{\bm{x}\in X^{\size{\sigma}}} \left[
                          \alloc(\bm{x}, \sigma)^{-x_i} \in V\right]
                          \iid_{\size{\sigma}}(\diff \bm{x}) \\
        &=   \sum_{i = 1}^{\size{\sigma}}  \frac{n_i}{K + 1}
                          \int_{\bm{x}\in X^{\size{\sigma^{n_i-}}}} \left[
                          \alloc(\bm{x}, \sigma^{n_i-}) \in V\right]
          \iid_{\size{\sigma}}(\diff \bm{x})\\
          &=   \sum_{n \in \supp(\sigma)}  \frac{n \cdot \sigma(n)}{K + 1}
                          \int_{\bm{x}\in X^{\size{\sigma^{n-}}}} \left[
                          \alloc(\bm{x}, \sigma^{n-}) \in V \right]
          \iid_{\size{\sigma}}(\diff \bm{x})
\end{align*}
which equals $(\base_\mu \circ \pdd)(\sigma)(V)$.
\end{proof}

\subsection{Main theorem}

We consider the diagram of spaces of partitions in
$\Kl(\G)$ 
\[
 \begin{tikzcd}
\Part(1) & \arrow[swap, ker]{l}{\pdd} \Part(2) & \arrow[swap, ker]{l}{\pdd} \cdots &
\arrow[swap, ker]{l}{} \Part(n) & \cdots \arrow[swap, ker]{l}{\pdd} 
\end{tikzcd}
\]
consisting of the draw-delete kernels from Section~\ref{sec:draw-delete-maps}.  Our main theorem is as follows:
\begin{theorem}[Kingman, categorical form]
  \label{thm:kingman}
The diagram 
\begin{equation}
  \label{eq:2}
\begin{tikzcd}
 &&&\elemfree \arrow[ker, bend right=10]{dll}{\pmn_2} 
 \arrow[ker, bend right=10, swap]{dlll}{\pmn_1}\arrow[ker]{d}{\pmn_n}\arrow[ker, bend left=10]{dr}{\pmn_{n +1}} &&       \\ 
  \Part(1) & \arrow[ker]{l}{\pdd} \Part(2) & \arrow[ker]{l}{\pdd} \cdots &
\arrow[swap, ker]{l}{} \Part(n) & \Part(n + 1) \arrow[ker]{l}{\pdd}
& \cdots \arrow[ker]{l}{\pdd} 
\end{tikzcd}
\end{equation}
is commutative, and the kernels $\{ \pmn_n : \elemfree \kerto \Part(n) \}$
form a limiting cone for the diagram of draw-delete maps in the
category $\Kl(G)$. 
\end{theorem}

As a consequence of this result, a consistent family of random
partitions corresponds to a unique random distributions.
Note that this theorem is stronger than Kingman's original result,
since we could be considering a measurable family of consistent
families (a cone over the diagram at an arbitrary object).

We will obtain this theorem as a corollary of its counterpart in
traditional probability theory, with elements. This is a
classical representation theorem for infinite exchangeable sequences due to de
Finetti \cite{de1929function}, and often regarded as a foundational result for Bayesian statistics. The categorical presentation of de Finetti's
theorem is due to Jacobs and Staton
\cite{sam-bart, samnote}.
\begin{theorem}[de Finetti, categorical form]
  \label{thm:definetti}
  Let $X$ be a standard Borel space. The multinomial kernels $\mn_n : \G X \kerto
  \M[n](X)$ form a limiting cone for the diagram of draw-delete
  maps in the category $\Kl(\G)$:
    \[
\begin{tikzcd}[column sep=2em]
 &&&\G X 
 \arrow[swap, ker, bend right=10]{dlll}{\mn_1}\arrow[ker, bend
 right=5]{dll}{\mn_2}\arrow[ker]{d}{\mn_{n }} \arrow[ker, bend left=10]{dr}{} &&       \\ 
  \M[1](X) &  \arrow[ker]{l}{\dd} \M[2](X) &
  \arrow[ker]{l}{\dd} \cdots &
\arrow[swap, ker]{l}{} \M[n](X) & \cdots \arrow[ker]{l}{\dd} 
\end{tikzcd}
\]
\end{theorem}

Our proof of Theorem~\ref{thm:kingman} uses the following categorical
fact\footnote{A proof is available at
  \url{https://ncatlab.org/nlab/show/retract\#RetractsOfDiagrams} .}:
\newcommand{\catI}{\mathbf{I}}
\newcommand{\catC}{\mathbf{C}}
\begin{lemma}
  \label{lem:categorical-fact}
  Let $\catI$ be a small category, and let $\catI^\triangleleft$ be
  the result of adding a free initial object to $\catI$.
  \begin{itemize}
    \item A functor $F : \catI^\triangleleft \to \catC$ is the same
      thing as a
  cone over the $\catI$-shaped diagram
  $\catI \hookrightarrow \catI^\triangleleft \xrightarrow{F} \catC$ in
  $\catC$.
\item Suppose that $F, G : \catI^\triangleleft \to \catC$ are functors
  such that  $G$  is a retract of $F$ in the functor category
  $[\catI^\triangleleft, \catC]$, \ie~we have natural transformations
  $s : F \to G$ and $t : G \to F$ such that $t \circ s = \id_G$. Then,
  if $F$ is a limit cone, so is $G$. 
\end{itemize}
\end{lemma}

Our version of Kingman's theorem follows from this fact, because the cone over
partitions is a retract of a de Finetti cone over multisets. 
\begin{proof}[Proof of Theorem~\ref{thm:kingman}]
  We instantiate Theorem~\ref{thm:definetti} at $X = [0,1]$ which
  gives a limit cone consisting of kernels
  $\mn_n : \G[0, 1] \kerto \M[n]([0, 1])$.  The rest of the proof
  follows from the series of results in this paper, showing that the
  families of maps $\base_{\uniform}$ and $\mc$, either between
  partitions and multisets or between distributions and element-free
  distributions, form section-retraction pairs which appropriately
  commute with $\dd$/$\pdd$ and $\mn$/$\pmn$. 
  We deduce that the diagram in \eqref{eq:2} commutes since for each $K$, 
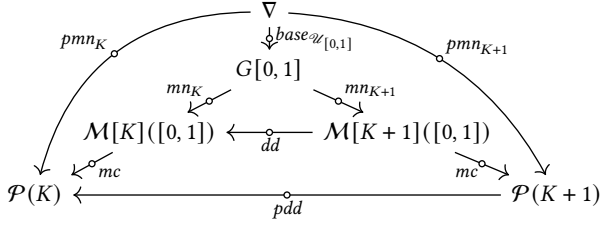
 \[
   \begin{tikzcd}[column sep=0em, row sep=1em]
	&& \elemfree \\
	&& {\G[0, 1]} \\
	& {\M[K]([0, 1])} && {\M[K+1]([0, 1])} \\
	{\Part(K)} &&&& {\Part(K +1)}
	\arrow["{\pmn_{K +1}}", ker, curve={height=-25pt}, from=1-3, to=4-5]
	\arrow["\pdd", ker, from=4-5, to=4-1]
	\arrow["{\pmn_K}"', ker, curve={height=30pt}, from=1-3, to=4-1]
	\arrow["{\base_{\uniform}}",ker,  from=1-3, to=2-3]
	\arrow["{\mn_K}"', ker, from=2-3, to=3-2]
	\arrow["\dd",ker, from=3-4, to=3-2]
	\arrow["\mc"', ker, from=3-4, to=4-5]
	\arrow["\mc", ker, from=3-2, to=4-1]
	\arrow["{\mn_{K +1}}", ker, from=2-3, to=3-4]
      \end{tikzcd}\]
    commutes. This shows that the maps $\pmn_K$ form a cone over the
    diagram of sets of partitions. A similar argument shows that
    we have a section-retraction pair in the functor
    category $[(\N^{\mathrm{op}})^\triangleleft, \Kl(G)]$, where $\N$ is
    the category $\{ 1 \to 2 \to \cdots \}$.
    We conclude by Lemma~\ref{lem:categorical-fact}.
\end{proof}

\section{Random element-free distributions as natural transformations}
\label{sec:random-element-free}

In this section we show how to arrive at element-free distributions from another
angle. Recall that, for every standard Borel space $X$, we have a
kernel $\base_\mu : \nabla \kerto \G X$
(Definition~\ref{def:base-kernel}). We observe that the same definition
actually gives a kernel
\[
   \base_X : \G X \times \nabla \kerto \G X
 \]
where the base measure $\mu$ is  regarded as an additional
argument. 
 
Given a distribution $\omega \in \G \elemfree$ on element-free distributions
(that is, $\omega : \One \kerto \elemfree$),
we construct a kernel denoted $(\omega \gg\!= \base_X)$ as follows:
\[
  \begin{tikzcd}
\G X \cong  \G X \times \One \arrow[ker]{r}{\G X \times \omega} &
 \G X \times \elemfree \arrow[ker]{r}{\base_X} & \G X
  \end{tikzcd}
\]
This kernel defines a natural transformation:

\begin{proposition}
  \label{prop:random-natural}
For $\omega \in \G \elemfree$, the family of measurable
functions $\omega \gg\!= \base_X : \G X \to \G \G X$, for $X \in
\Sbs$, defines a natural transformation between functors $\Sbs \to
\Sbs$. More concretely, for every measurable function $f : X \to Y$, the
following diagram commutes:
\[
  \begin{tikzcd}[column sep=5em]
    \G X \arrow{r}{\omega \gg\!= \base_X} \arrow[swap]{d}{\G f} & \G
    \G X \arrow{d}{\G \G f}\\
    \G Y \arrow[swap]{r}{\omega \gg\!= \base_Y} & \G \G Y
  \end{tikzcd}
\]
\end{proposition}  
\begin{proof}
  Although we stated the lemma in this way for the purposes of the
  section below, this follows easily from the more primitive fact that
  each $\base_X(-, \varphi)$ is natural, which in turn is an easy
  consequence of the naturality of $\iid_{\size{\varphi}}$. 
\end{proof}

\newcommand{\Nat}{\mathrm{Nat}} Proposition~\ref{prop:random-natural}
defines a function $\G \elemfree \to \Nat(\G, \G\G)$, where
$\Nat(\G, \G\G)$ is the set of all natural transformations.  The main
theorem of this section states that this function is bijective, \ie~
that every natural transformation $\G \to \G\G$ is of the form
$\omega \gg\!= \base$ for a unique $\omega \in \G \elemfree$. 

Our proof relies on the following well-known property of standard Borel
spaces. (This is sometimes known as the randomization lemma, \eg~\cite[Lemma~3.22]{fmp}.)
\begin{lemma}
  \label{lem:randomization}
  For every standard Borel space $X$ and probability measure $\mu \in \G
X$, there exists $f : [0, 1] \to X$ such that $\mu = G(f)(\uniform)$.
\end{lemma}

We deduce that a natural transformation $\G \to
\G\G$ is determined by its action on the uniform distribution
$\uniform$. 
\begin{lemma}
  \label{lem:equality_nat_trans}
For natural transformations $H, H' : \G \to \G \G$, if $H_{[0, 1]}(\uniform) = H'_{[0, 1]}(\uniform)$, then $H = H'$. 
\end{lemma}
\begin{proof}
Let $X \in \Sbs$ and $\mu \in \G X$. We show that $H_X(\mu) =
H'_X(\mu)$. By Lemma~\ref{lem:randomization}, there is some $f : [0, 1] \to X$ such
that $\mu = \G(f)(\uniform)$. By naturality, we have that
\[H_X(\mu) = H_X(\G(f)(\uniform)) =\G\G(f)(H_{[0, 1]}(\uniform)),\] and similarly for
$H'_X$. Thus 
\[
H_X(\mu) = \G\G(f)(H_{[0, 1]}(\uniform)) = \G\G(f)(H'_{[0,
  1]}(\uniform)) = H'_X(\mu)
\]
and we are done.
\end{proof}

\begin{theorem}
  \label{thm:correspondence}
There is an isomorphism $\G\elemfree \cong \Nat(\G, \G\G)$, given by
the pair of inverse maps $\alpha$ and $\beta$ given below. 
  \begin{align*}
  &\alpha\  :\ \G \elemfree \longrightarrow \Nat(\G, \G\G)  \ : \
  \omega \longmapsto \{ \omega \gg\!= \base_X \}_{X \in \Sbs} \\
  &\beta\ :\  \Nat(\G, \G\G) \longrightarrow \G \elemfree  \ : \ 
   H \longmapsto \G(\mc)(H_{[0, 1]}(\uniform)) 
\end{align*}
\end{theorem}
\begin{proof}
  The fact that $\beta \circ \alpha = \id_{\G \elemfree}$ follows
  essentially from the retract property of
  Lemma~\ref{prop:uniform-base}, to which we apply $\G$. We show 
  $\alpha \circ \beta = \id_{\Nat(\G, \G\G)}$, for which it suffices
  to show that $\beta$ is injective. So we suppose $H, H' \in
  \Nat(\G, \G\G)$ satisfy $\G(\mc)(H_{[0, 1]}(\uniform)) = \G(\mc)(H'_{[0, 1]}(\uniform))$.
 
  To show $H = H'$, it suffices to show that $H_{[0, 1]}(\uniform) =
  H'_{[0, 1]}(\uniform) \in \G\G[0, 1]$,  by
  Lemma~\ref{lem:equality_nat_trans}. By the \emph{uniqueness} part
  of de Finetti's theorem (Theorem~\ref{thm:definetti}), 
  it suffices to show that the family of measures 
  \[
    \begin{tikzcd}
\mu_K \arrow[phantom]{r}[description]{=}&      1 \arrow[ker]{r}{H_{[0, 1]}(\uniform)} &[3em] \G[0, 1]
      \arrow[ker]{r}{\mn_K} & \M[K]([0, 1]),
    \end{tikzcd}
  \]
  for $K \geq 0$, coincides with the family of measures $\mu_K'$
  induced by $H'$ in the same way. So we now fix $K \geq 1$ and show
  $\mu_K' = \mu_K$. First we observe that $\G(\mc)(\mu_K') =
  \G(\mc)(\mu_K)$, by an easy argument using that
  $\G(\mc)(H_{[0, 1]}(\uniform)) =  \G(\mc)(H'_{[0, 1]}(\uniform))$. 

  Now, using the naturality of $H$ and $\mn$, we can show that
  $\mu_K$ and $\mu'_K$ are invariant under any
  uniform-measure-preserving function $\phi : [0, 1] \to [0, 1]$.  The
  full argument (for $\mu_K$) is as follows: 
    
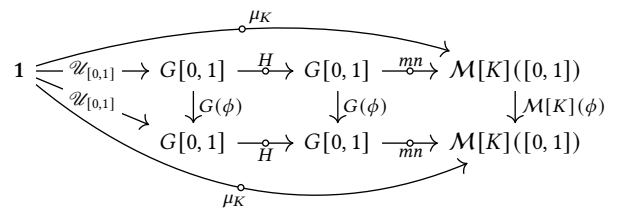
\[\begin{tikzcd}[row sep=1.4em]
	\One && {\G[0, 1]} & {\G[0, 1]} & {\M[K]([0, 1])} \\
	&& {\G[0, 1]} & {\G[0, 1]} & {\M[K]([0, 1])}
	\arrow["\uniform"{description}, from=1-1, to=1-3]
	\arrow["{\mu_K}", curve={height=-20pt}, from=1-1, to=1-5, ker]
	\arrow["\uniform"{description}, from=1-1, to=2-3]
	\arrow["{\mu_K}"', curve={height=40pt}, from=1-1, to=2-5, ker]
	\arrow["H", from=1-3, to=1-4, ker]
	\arrow["{\G(\phi)}", from=1-3, to=2-3]
	\arrow["\mn", from=1-4, to=1-5, ker]
	\arrow["{\G(\phi)}", from=1-4, to=2-4]
	\arrow["{\M[K](\phi)}", from=1-5, to=2-5]
	\arrow["{H}"', from=2-3, to=2-4, ker]
	\arrow["\mn"', from=2-4, to=2-5, ker]
\end{tikzcd}\]
Applying Lemma~\ref{lem:random-element-free}, and the fact that
$\G(\mc)(\mu_K) = \G(\mc)(\mu_K')$, we conclude that $\mu_K = \mu_K'$.
\end{proof}

The technical lemma below completes the proof. 
\begin{lemma}
  \label{lem:random-element-free}
Let $\mu$ be a distribution on $\M[K]([0, 1])$ such that
$G(\M[K](\phi))(\mu) = \mu$ for every uniform-measure-preserving function
$\phi : [0, 1] \to [0, 1]$. Then the following diagram commutes:
\[
\begin{tikzcd}
	\One &&& {\M[K]([0, 1])} \\
	& {\M[K]([0, 1])} & {\Part(K)}
	\arrow["\mu", from=1-1, to=1-4, ker]
	\arrow["\mu"', from=1-1, to=2-2, ker]
	\arrow["\mc"', from=2-2, to=2-3]
	\arrow["{\base_{\uniform}}"', from=2-3, to=1-4, ker]
\end{tikzcd}
\]
\end{lemma}
\newcommand{\Cube}{\mathcal{C}}
\begin{proof}[Proof sketch]
It is enough to show that $\mu$ is completely determined by the distribution
$\G(\mc)(\mu)$ on $\Part(K)$. We consider the disjoint union $\M[K](0, 1)
= \bigcup_{\sigma \in \Part(K)} \mc^{-1}\{\sigma\}$, and we
approximate each $\mc^{-1}\{ \sigma\}$ by an increasing chain, as
follows. Let $B^n_m \subseteq [0, 1]$ be the approximating intervals
defined in the proof of Lemma~\ref{lem:epsilon}, for $m \in \N$ and $0
\leq n < 2^m$. We can enumerate the blocks of any $\sigma
\in \Part(K)$ as $k_1 \leq \dots \leq k_{\size{\sigma}}$. Then if $m
\in \N$, and $\boldsymbol{n} = (n_1, \dots, n_{\size{\sigma}})$ with
all $n_i < 2^m$ and \emph{distinct}, define
  \[
\Cube(\sigma, m, \boldsymbol{n}) = \left\{ \varphi \in \M[K]([0, 1]) \mid
\varphi = \sum_{i = 1}^{\size{\sigma}} k_i[x_i], \text{ for } x_i \in B_m^{n_i} \right\}. 
\]
We show that, when $\boldsymbol{n}$ ranges over tuples as
above, 
\[
 \mc^{-1} \{ \sigma \} = \bigcup_{m = 1}^\infty
 \bigcup_{\boldsymbol{n}} \Cube(\sigma, m, \boldsymbol{n}).
\]
The invariance under any uniform-measure-preserving $\phi$ is then used
to derive that $\mu(\Cube(\sigma, m, \boldsymbol{n}))$ does not
depend on $\boldsymbol{n}$. By an additional limit argument we 
then deduce that $\mu(\Cube(\sigma, m, \boldsymbol{n}))$ is a constant
fraction of $\mu(\mc(\sigma))$, and so in particular the value of
$\mu$ on these sets is determined by $\G(\mc)(\mu)$. We are done, because the sets $\Cube(\sigma,
m, \boldsymbol{n})$ form a basis for $\Sigma_{\M[K]([0, 1])}$. 
\end{proof}

\section{Conclusions}
\label{sec:conclusion}

\paragraph{Summary}
We have given new definitions and tools for the element-free
distributions of Kingman \cite{kingman-representation} in
the study of partition structures. Our formalization makes heavy use
of multisets, showing that the coefficients of an element-free distribution
do not need to be ordered. 

We have studied the two key operations relating
ordinary and element-free probability: multiplicity count,
which forgets elements, and the operation of drawing new elements from
a base measure. The latter seems new to this paper.

Our development uses technical measure-theoretic tools, but we
emphasize that including continuous
distributions is crucial for our main representation theorems to
hold. (For an edge-case example, the infinite random partition which
has all singletons with probability 1 can only be represented by a nonatomic distribution.)

\paragraph{Related work}
A major inspiration for this paper is the line of research by Jacobs on structural
probability theory
(\eg~\cite{jacobs2022reconstruction,Jacobs_2022,jacobs2022stick}). The
idea of multiset-based element-free probability is due to
him, and we have referred to many of his results on partitions
\cite{lics-jacobs}. Our version of Kingman's theorem is the answer to
an open problem in \cite[\S12]{lics-jacobs}, and our proof relies on
key insights about probabilistic representation theorems due to Jacobs
and Staton \cite{sam-bart,samnote}. For multisets over standard Borel spaces,
another key inspiration is the monad for point processes of Dash and Staton \cite{DBLP:journals/corr/abs-2101-10479}.

In another line of research, Danos and Garnier have observed \cite{dirichlet-is-natural} that the
Dirichlet Process determines a natural transformation. This inspired the
work of Section~\ref{sec:random-element-free}, and our Theorem~\ref{thm:correspondence}
should recover the stick-breaking weights for the Dirichlet Process, in the
form of an element-free distribution.

\paragraph{Perspectives}
There are several avenues for further work. 

Nonparametric Bayesian processes influenced the early development of
probabilistic programming \cite{church}. More recently typed languages
with polymorphism were designed for probabilistic modelling
\cite{lazyppl,scibior2015practical}, and it seems clear that a
polymorphic implementation of the Dirichlet process must treat the
atoms and the element-free part independently. Thus an interface
for element-free distributions would be convenient. A
theoretical investigation relating naturality and polymorphism in a
probabilistic setting (in the style of \cite{wadler1989theorems})
seems important for proper foundations of nonparametric probabilistic languages.

Beyond element-free distributions, we can ask whether similar
constructions and theorems exist for arbitrary element-free \emph{unnormalized}
measures. Unnormalized measures are fundamental in probabilistic programming
\cite{staton2017commutative}, and random measures play a key role
in nonparametric statistics (\eg~as point processes \cite{last-penrose} or as latent
feature models \cite{roy2014continuum, thibaux2007hierarchical}),
also involving base measures. 

We finally note that multisets and probabilities have been combined in the
context of probabilistic models of linear logic
(\eg~\cite{danos2011probabilistic}). A deeper connection to
investigate is the construction of the free linear exponential
(\eg~\cite{crubille2017free, dahlqvist2019semantics})
using a universal categorical construction over diagrams of
measures, similar to 
Theorem~\ref{thm:kingman} and Theorem~\ref{thm:definetti}.

\section*{Acknowledgements}
We are grateful for helpful conversations with Sam Staton, Ned
Summers, Paolo Perrone, and Ohad Kammar. The results in
\S\ref{sec:random-element-free} were initially presented at LAFI 2023,
where we received useful feedback from reviewers and other
attendees. Thanks to the anonymous LICS reviewers.
        
We acknowledge support from a Royal Society University Research
Fellowship, ERC grant BLaSt, and a Paris Region Fellowship
co-funded by the European Union (Marie Sk\l{}odowska-Curie grant
agreement 945298).

\bibliographystyle{ACM-Reference-Format}
 \bibliography{paper}


\begin{thebibliography}{36}

%%% the Web address.

\ifx \showCODEN    \undefined \def \showCODEN     #1{\unskip}     \fi
\ifx \showDOI      \undefined \def \showDOI       #1{#1}\fi
\ifx \showISBNx    \undefined \def \showISBNx     #1{\unskip}     \fi
\ifx \showISBNxiii \undefined \def \showISBNxiii  #1{\unskip}     \fi
\ifx \showISSN     \undefined \def \showISSN      #1{\unskip}     \fi
\ifx \showLCCN     \undefined \def \showLCCN      #1{\unskip}     \fi
\ifx \shownote     \undefined \def \shownote      #1{#1}          \fi
\ifx \showarticletitle \undefined \def \showarticletitle #1{#1}   \fi
\ifx \showURL      \undefined \def \showURL       {\relax}        \fi
% invisible to TeX
\providecommand\bibfield[2]{#2}
\providecommand\bibinfo[2]{#2}
\providecommand\natexlab[1]{#1}
\providecommand\showeprint[2][]{arXiv:#2}

\bibitem[Aldous(1985)]%
        {10.1007/BFb0099421}
\bibfield{author}{\bibinfo{person}{David~J. Aldous}.}
  \bibinfo{year}{1985}\natexlab{}.
\newblock \showarticletitle{Exchangeability and related topics}. In
  \bibinfo{booktitle}{\emph{{\'E}cole d'{\'E}t{\'e} de Probabilit{\'e}s de
  Saint-Flour XIII --- 1983}}. \bibinfo{publisher}{Springer Berlin Heidelberg},
  \bibinfo{address}{Berlin, Heidelberg}, \bibinfo{pages}{1--198}.
\newblock


\bibitem[Andrews(1998)]%
        {andrews1998theory}
\bibfield{author}{\bibinfo{person}{George~E Andrews}.}
  \bibinfo{year}{1998}\natexlab{}.
\newblock \bibinfo{booktitle}{\emph{The theory of partitions}}.
\newblock Number~2. \bibinfo{publisher}{Cambridge university press}.
\newblock


\bibitem[\c~Cinlar(2011)]%
        {cinlar2011probability}
\bibfield{author}{\bibinfo{person}{Erhan \c Cinlar}.}
  \bibinfo{year}{2011}\natexlab{}.
\newblock \bibinfo{booktitle}{\emph{Probability and stochastics}}.
  Vol.~\bibinfo{volume}{261}.
\newblock \bibinfo{publisher}{Springer}.
\newblock


\bibitem[Crubill{\'e} et~al\mbox{.}(2017)]%
        {crubille2017free}
\bibfield{author}{\bibinfo{person}{Rapha{\"e}lle Crubill{\'e}},
  \bibinfo{person}{Thomas Ehrhard}, \bibinfo{person}{Michele Pagani}, {and}
  \bibinfo{person}{Christine Tasson}.} \bibinfo{year}{2017}\natexlab{}.
\newblock \showarticletitle{The free exponential modality of probabilistic
  coherence spaces}. In \bibinfo{booktitle}{\emph{International Conference on
  Foundations of Software Science and Computation Structures}}. Springer,
  \bibinfo{pages}{20--35}.
\newblock


\bibitem[Dahlqvist and Kozen(2019)]%
        {dahlqvist2019semantics}
\bibfield{author}{\bibinfo{person}{Fredrik Dahlqvist} {and}
  \bibinfo{person}{Dexter Kozen}.} \bibinfo{year}{2019}\natexlab{}.
\newblock \showarticletitle{Semantics of higher-order probabilistic programs
  with conditioning}.
\newblock \bibinfo{journal}{\emph{Proceedings of the ACM on Programming
  Languages}} \bibinfo{volume}{4}, \bibinfo{number}{POPL}
  (\bibinfo{year}{2019}), \bibinfo{pages}{1--29}.
\newblock


\bibitem[Danos and Ehrhard(2011)]%
        {danos2011probabilistic}
\bibfield{author}{\bibinfo{person}{Vincent Danos} {and} \bibinfo{person}{Thomas
  Ehrhard}.} \bibinfo{year}{2011}\natexlab{}.
\newblock \showarticletitle{Probabilistic coherence spaces as a model of
  higher-order probabilistic computation}.
\newblock \bibinfo{journal}{\emph{Information and Computation}}
  \bibinfo{volume}{209}, \bibinfo{number}{6} (\bibinfo{year}{2011}),
  \bibinfo{pages}{966--991}.
\newblock


\bibitem[Danos and Garnier(2015)]%
        {dirichlet-is-natural}
\bibfield{author}{\bibinfo{person}{Vincent Danos} {and} \bibinfo{person}{Ilias
  Garnier}.} \bibinfo{year}{2015}\natexlab{}.
\newblock \showarticletitle{Dirichlet is natural}.
\newblock \bibinfo{journal}{\emph{Electronic Notes in Theoretical Computer
  Science}}  \bibinfo{volume}{319} (\bibinfo{year}{2015}),
  \bibinfo{pages}{137--164}.
\newblock


\bibitem[Dash et~al\mbox{.}(2023)]%
        {lazyppl}
\bibfield{author}{\bibinfo{person}{Swaraj Dash}, \bibinfo{person}{Younesse
  Kaddar}, \bibinfo{person}{Hugo Paquet}, {and} \bibinfo{person}{Sam Staton}.}
  \bibinfo{year}{2023}\natexlab{}.
\newblock \showarticletitle{Affine Monads and Lazy Structures for Bayesian
  Programming}.
\newblock \bibinfo{journal}{\emph{Proc. {ACM} Program. Lang.}}
  \bibinfo{volume}{7}, \bibinfo{number}{{POPL}} (\bibinfo{year}{2023}),
  \bibinfo{pages}{1338--1368}.
\newblock
\urldef\tempurl%
\url{https://doi.org/10.1145/3571239}
\showDOI{\tempurl}


\bibitem[Dash and Staton(2020)]%
        {DBLP:journals/corr/abs-2101-10479}
\bibfield{author}{\bibinfo{person}{Swaraj Dash} {and} \bibinfo{person}{Sam
  Staton}.} \bibinfo{year}{2020}\natexlab{}.
\newblock \showarticletitle{A Monad for Probabilistic Point Processes}. In
  \bibinfo{booktitle}{\emph{Proceedings of the 3rd Annual International Applied
  Category Theory Conference 2020, {ACT} 2020, Cambridge, USA, 6-10th July
  2020}} \emph{(\bibinfo{series}{{EPTCS}}, Vol.~\bibinfo{volume}{333})},
  \bibfield{editor}{\bibinfo{person}{David~I. Spivak} {and}
  \bibinfo{person}{Jamie Vicary}} (Eds.). \bibinfo{pages}{19--32}.
\newblock
\urldef\tempurl%
\url{https://doi.org/10.4204/EPTCS.333.2}
\showDOI{\tempurl}


\bibitem[Dash and Staton(2021)]%
        {DBLP:journals/corr/abs-2112-14048}
\bibfield{author}{\bibinfo{person}{Swaraj Dash} {and} \bibinfo{person}{Sam
  Staton}.} \bibinfo{year}{2021}\natexlab{}.
\newblock \showarticletitle{Monads for Measurable Queries in Probabilistic
  Databases}. In \bibinfo{booktitle}{\emph{Proceedings 37th Conference on
  Mathematical Foundations of Programming Semantics, {MFPS} 2021, Hybrid:
  Salzburg, Austria and Online, 30th August - 2nd September, 2021}}
  \emph{(\bibinfo{series}{{EPTCS}}, Vol.~\bibinfo{volume}{351})},
  \bibfield{editor}{\bibinfo{person}{Ana Sokolova}} (Ed.).
  \bibinfo{pages}{34--50}.
\newblock
\urldef\tempurl%
\url{https://doi.org/10.4204/EPTCS.351.3}
\showDOI{\tempurl}


\bibitem[De~Finetti(1929)]%
        {de1929function}
\bibfield{author}{\bibinfo{person}{Bruno De~Finetti}.}
  \bibinfo{year}{1929}\natexlab{}.
\newblock \showarticletitle{Characteristic function of a random phenomenon}. In
  \bibinfo{booktitle}{\emph{Proceedings of the International Congress of
  Mathematicians: Bologna from 3 to 10 September 1928}}.
  \bibinfo{pages}{179--190}.
\newblock


\bibitem[Ferguson(1973)]%
        {ferguson1973bayesian}
\bibfield{author}{\bibinfo{person}{Thomas~S Ferguson}.}
  \bibinfo{year}{1973}\natexlab{}.
\newblock \showarticletitle{A Bayesian analysis of some nonparametric
  problems}.
\newblock \bibinfo{journal}{\emph{The annals of statistics}}
  (\bibinfo{year}{1973}), \bibinfo{pages}{209--230}.
\newblock


\bibitem[Fritz et~al\mbox{.}(2021)]%
        {fritz2021finetti}
\bibfield{author}{\bibinfo{person}{Tobias Fritz},
  \bibinfo{person}{Tom{\'a}{\v{s}} Gonda}, {and} \bibinfo{person}{Paolo
  Perrone}.} \bibinfo{year}{2021}\natexlab{}.
\newblock \showarticletitle{De {F}inetti’s Theorem in Categorical
  Probability}.
\newblock \bibinfo{journal}{\emph{Journal of Stochastic Analysis}}
  \bibinfo{volume}{2}, \bibinfo{number}{4} (\bibinfo{year}{2021}),
  \bibinfo{pages}{6}.
\newblock


\bibitem[Gelman et~al\mbox{.}(1995)]%
        {gelman1995bayesian}
\bibfield{author}{\bibinfo{person}{Andrew Gelman}, \bibinfo{person}{John~B
  Carlin}, \bibinfo{person}{Hal~S Stern}, {and} \bibinfo{person}{Donald~B
  Rubin}.} \bibinfo{year}{1995}\natexlab{}.
\newblock \bibinfo{booktitle}{\emph{Bayesian data analysis}}.
\newblock \bibinfo{publisher}{Chapman and Hall/CRC}.
\newblock


\bibitem[Giry(1982)]%
        {giry}
\bibfield{author}{\bibinfo{person}{Michele Giry}.}
  \bibinfo{year}{1982}\natexlab{}.
\newblock \showarticletitle{A categorical approach to probability theory}.
\newblock In \bibinfo{booktitle}{\emph{Categorical aspects of topology and
  analysis}}. \bibinfo{publisher}{Springer}, \bibinfo{pages}{68--85}.
\newblock


\bibitem[Goodman et~al\mbox{.}(2008)]%
        {church}
\bibfield{author}{\bibinfo{person}{Noah~D Goodman}, \bibinfo{person}{Vikash~K
  Mansinghka}, \bibinfo{person}{Daniel Roy}, \bibinfo{person}{Keith Bonawitz},
  {and} \bibinfo{person}{Joshua~B Tenenbaum}.} \bibinfo{year}{2008}\natexlab{}.
\newblock \showarticletitle{Church: a language for generative models}. In
  \bibinfo{booktitle}{\emph{Proceedings of the Twenty-Fourth Conference on
  Uncertainty in Artificial Intelligence}}. \bibinfo{pages}{220--229}.
\newblock


\bibitem[Jacobs(2022a)]%
        {jacobsone}
\bibfield{author}{\bibinfo{person}{Bart Jacobs}.}
  \bibinfo{year}{2022}\natexlab{a}.
\newblock \bibinfo{title}{Basic Combinatorics and Sufficient Statistics for
  Mutations on Multiple Datatypes}.
\newblock
\newblock
\urldef\tempurl%
\url{https://researchers.one/articles/22.11.00003v1}
\showURL{%
\tempurl}
\newblock
\shownote{Researchers.One}.


\bibitem[Jacobs(2022b)]%
        {lics-jacobs}
\bibfield{author}{\bibinfo{person}{Bart Jacobs}.}
  \bibinfo{year}{2022}\natexlab{b}.
\newblock \showarticletitle{Partitions and {E}wens distributions in
  element-free probability theory}. In \bibinfo{booktitle}{\emph{Proceedings of
  the 37th Annual ACM/IEEE Symposium on Logic in Computer Science}}.
  \bibinfo{pages}{1--9}.
\newblock


\bibitem[Jacobs(2022c)]%
        {jacobs2022reconstruction}
\bibfield{author}{\bibinfo{person}{Bart Jacobs}.}
  \bibinfo{year}{2022}\natexlab{c}.
\newblock \showarticletitle{A Reconstruction of Ewens’ Sampling Formula via
  Lists of Coins}.
\newblock In \bibinfo{booktitle}{\emph{A Journey from Process Algebra via Timed
  Automata to Model Learning: Essays Dedicated to Frits Vaandrager on the
  Occasion of His 60th Birthday}}. \bibinfo{publisher}{Springer},
  \bibinfo{pages}{339--357}.
\newblock


\bibitem[Jacobs(2022d)]%
        {jacobs2022stick}
\bibfield{author}{\bibinfo{person}{Bart Jacobs}.}
  \bibinfo{year}{2022}\natexlab{d}.
\newblock \showarticletitle{Stick breaking, in coalgebra and probability}. In
  \bibinfo{booktitle}{\emph{International Workshop on Coalgebraic Methods in
  Computer Science}}. Springer, \bibinfo{pages}{176--193}.
\newblock


\bibitem[Jacobs(2022e)]%
        {Jacobs_2022}
\bibfield{author}{\bibinfo{person}{Bart Jacobs}.}
  \bibinfo{year}{2022}\natexlab{e}.
\newblock \showarticletitle{Urns and Tubes}.
\newblock \bibinfo{journal}{\emph{Compositionality}}  \bibinfo{volume}{4}
  (\bibinfo{date}{Dec.} \bibinfo{year}{2022}), \bibinfo{pages}{4}.
\newblock
\showISSN{2631-4444}
\urldef\tempurl%
\url{https://doi.org/10.32408/compositionality-4-4}
\showDOI{\tempurl}


\bibitem[Jacobs and Staton(2020a)]%
        {sam-bart}
\bibfield{author}{\bibinfo{person}{Bart Jacobs} {and} \bibinfo{person}{Sam
  Staton}.} \bibinfo{year}{2020}\natexlab{a}.
\newblock \showarticletitle{De {F}inetti’s construction as a categorical
  limit}. In \bibinfo{booktitle}{\emph{International Workshop on Coalgebraic
  Methods in Computer Science}}. Springer, \bibinfo{pages}{90--111}.
\newblock


\bibitem[Jacobs and Staton(2020b)]%
        {samnote}
\bibfield{author}{\bibinfo{person}{Bart Jacobs} {and} \bibinfo{person}{Sam
  Staton}.} \bibinfo{year}{2020}\natexlab{b}.
\newblock \bibinfo{title}{Unpublished appendix to CMCS 2020: general case of de
  Finetti}.  (\bibinfo{year}{2020}).
\newblock


\bibitem[Jacobs and Stein(2023)]%
        {DBLP:conf/csl/0001S23}
\bibfield{author}{\bibinfo{person}{Bart Jacobs} {and} \bibinfo{person}{Dario
  Stein}.} \bibinfo{year}{2023}\natexlab{}.
\newblock \showarticletitle{Counting and Matching}. In
  \bibinfo{booktitle}{\emph{31st {EACSL} Annual Conference on Computer Science
  Logic, {CSL} 2023, February 13-16, 2023, Warsaw, Poland}}
  \emph{(\bibinfo{series}{LIPIcs}, Vol.~\bibinfo{volume}{252})},
  \bibfield{editor}{\bibinfo{person}{Bartek Klin} {and} \bibinfo{person}{Elaine
  Pimentel}} (Eds.). \bibinfo{publisher}{Schloss Dagstuhl - Leibniz-Zentrum
  f{\"{u}}r Informatik}, \bibinfo{pages}{28:1--28:15}.
\newblock
\urldef\tempurl%
\url{https://doi.org/10.4230/LIPICS.CSL.2023.28}
\showDOI{\tempurl}


\bibitem[Kallenberg(2021)]%
        {fmp}
\bibfield{author}{\bibinfo{person}{Olav Kallenberg}.}
  \bibinfo{year}{2021}\natexlab{}.
\newblock \bibinfo{booktitle}{\emph{Foundations of Modern Probability}
  (\bibinfo{edition}{hardcover} ed.)}.
\newblock \bibinfo{publisher}{Springer}. 958 pages.
\newblock


\bibitem[Kingman(1978)]%
        {kingman-representation}
\bibfield{author}{\bibinfo{person}{John~FC Kingman}.}
  \bibinfo{year}{1978}\natexlab{}.
\newblock \showarticletitle{The representation of partition structures}.
\newblock \bibinfo{journal}{\emph{Journal of the London Mathematical Society}}
  \bibinfo{volume}{2}, \bibinfo{number}{2} (\bibinfo{year}{1978}),
  \bibinfo{pages}{374--380}.
\newblock


\bibitem[Last and Penrose(2017)]%
        {last-penrose}
\bibfield{author}{\bibinfo{person}{G{\"u}nter Last} {and}
  \bibinfo{person}{Mathew Penrose}.} \bibinfo{year}{2017}\natexlab{}.
\newblock \bibinfo{booktitle}{\emph{Lectures on the {P}oisson process}}.
  Vol.~\bibinfo{volume}{7}.
\newblock \bibinfo{publisher}{Cambridge University Press}.
\newblock


\bibitem[Lawvere(1963)]%
        {lawvere1963functorial}
\bibfield{author}{\bibinfo{person}{F~William Lawvere}.}
  \bibinfo{year}{1963}\natexlab{}.
\newblock \showarticletitle{Functorial semantics of algebraic theories}.
\newblock \bibinfo{journal}{\emph{Proceedings of the National Academy of
  Sciences}} \bibinfo{volume}{50}, \bibinfo{number}{5} (\bibinfo{year}{1963}),
  \bibinfo{pages}{869--872}.
\newblock


\bibitem[Pitman(2006)]%
        {pitman2006combinatorial}
\bibfield{author}{\bibinfo{person}{Jim Pitman}.}
  \bibinfo{year}{2006}\natexlab{}.
\newblock \bibinfo{booktitle}{\emph{Combinatorial stochastic processes: Ecole
  d'et{\'e} de probabilit{\'e}s de saint-flour xxxii-2002}}.
\newblock \bibinfo{publisher}{Springer}.
\newblock


\bibitem[Pitman and Yor(1997)]%
        {pitman1997two}
\bibfield{author}{\bibinfo{person}{Jim Pitman} {and} \bibinfo{person}{Marc
  Yor}.} \bibinfo{year}{1997}\natexlab{}.
\newblock \showarticletitle{The two-parameter Poisson-Dirichlet distribution
  derived from a stable subordinator}.
\newblock \bibinfo{journal}{\emph{The Annals of Probability}}
  (\bibinfo{year}{1997}), \bibinfo{pages}{855--900}.
\newblock


\bibitem[Roy(2014)]%
        {roy2014continuum}
\bibfield{author}{\bibinfo{person}{Daniel~M Roy}.}
  \bibinfo{year}{2014}\natexlab{}.
\newblock \showarticletitle{The continuum-of-urns scheme, generalized beta and
  Indian buffet processes, and hierarchies thereof}.
\newblock \bibinfo{journal}{\emph{arXiv preprint arXiv:1501.00208}}
  (\bibinfo{year}{2014}).
\newblock


\bibitem[{\'S}cibior et~al\mbox{.}(2015)]%
        {scibior2015practical}
\bibfield{author}{\bibinfo{person}{Adam {\'S}cibior}, \bibinfo{person}{Zoubin
  Ghahramani}, {and} \bibinfo{person}{Andrew~D Gordon}.}
  \bibinfo{year}{2015}\natexlab{}.
\newblock \showarticletitle{Practical probabilistic programming with monads}.
  In \bibinfo{booktitle}{\emph{Proceedings of the 2015 ACM SIGPLAN Symposium on
  Haskell}}. \bibinfo{pages}{165--176}.
\newblock


\bibitem[Sethuraman(1994)]%
        {sethuraman1994constructive}
\bibfield{author}{\bibinfo{person}{Jayaram Sethuraman}.}
  \bibinfo{year}{1994}\natexlab{}.
\newblock \showarticletitle{A constructive definition of {D}irichlet priors}.
\newblock \bibinfo{journal}{\emph{Statistica sinica}} (\bibinfo{year}{1994}),
  \bibinfo{pages}{639--650}.
\newblock


\bibitem[Staton(2017)]%
        {staton2017commutative}
\bibfield{author}{\bibinfo{person}{Sam Staton}.}
  \bibinfo{year}{2017}\natexlab{}.
\newblock \showarticletitle{Commutative semantics for probabilistic
  programming}. In \bibinfo{booktitle}{\emph{Programming Languages and Systems:
  26th European Symposium on Programming, ESOP 2017, Held as Part of the
  European Joint Conferences on Theory and Practice of Software, ETAPS 2017,
  Uppsala, Sweden, April 22--29, 2017, Proceedings 26}}. Springer,
  \bibinfo{pages}{855--879}.
\newblock


\bibitem[Thibaux and Jordan(2007)]%
        {thibaux2007hierarchical}
\bibfield{author}{\bibinfo{person}{Romain Thibaux} {and}
  \bibinfo{person}{Michael~I Jordan}.} \bibinfo{year}{2007}\natexlab{}.
\newblock \showarticletitle{Hierarchical beta processes and the Indian buffet
  process}. In \bibinfo{booktitle}{\emph{Artificial intelligence and
  statistics}}. PMLR, \bibinfo{pages}{564--571}.
\newblock


\bibitem[Wadler(1989)]%
        {wadler1989theorems}
\bibfield{author}{\bibinfo{person}{Philip Wadler}.}
  \bibinfo{year}{1989}\natexlab{}.
\newblock \showarticletitle{Theorems for free!}. In
  \bibinfo{booktitle}{\emph{Proceedings of the fourth international conference
  on Functional programming languages and computer architecture}}.
  \bibinfo{pages}{347--359}.
\newblock


\end{thebibliography}

\end{document}